\documentclass[aps,pra,reprint, amsmath, amssymb,superscriptaddress]{revtex4-1}

\pdfoutput=1
\usepackage[utf8]{inputenc}
\usepackage{bm}
\usepackage[english]{babel}
\usepackage[T1]{fontenc}
\usepackage{mathrsfs}
\usepackage[retainorgcmds]{IEEEtrantools}
\usepackage{amsmath}
\usepackage{amssymb}
\usepackage{color}
\usepackage{amsfonts}
\usepackage{times,txfonts}
\usepackage{nicefrac}
\usepackage[colorlinks=true,linkcolor=blue,urlcolor=blue,citecolor=blue,pdfusetitle]{hyperref}
\usepackage{float}
\usepackage{epsfig} 
\usepackage{subfigure}
\usepackage{physics}
\usepackage{amssymb}
\usepackage[amssymb]{SIunits}
\graphicspath{{Images/}}
\usepackage{xcolor}

\newcommand{\mean}[1]{\ensuremath{\langle #1 \rangle}}
\newcommand{\bigpar}[1]{\ensuremath{\left( #1 \right)}}
\newcommand{\dg}[1]{\ensuremath{#1^{\dagger}}}
\newcommand{\cj}[1]{\ensuremath{\bar{#1}}}
\newcommand{\pd}[2]{\ensuremath{\partial_{#1} #2}} 
\newcommand{\dpd}[2]{\ensuremath{\partial_{#1}^2 #2}} 

\graphicspath{{Images/}}

\begin{document}

\title{Wehrl entropy production rate across a dynamical quantum phase transition}

\author{B. O. Goes}
\affiliation{Instituto de F\'{i}sica, Universidade de S\~{a}o Paulo, CEP 05314-970, S\~{a}o Paulo, S\~{a}o Paulo, Brazil}

\author{G. T. Landi}
\affiliation{Instituto de F\'{i}sica, Universidade de S\~{a}o Paulo, CEP 05314-970, S\~{a}o Paulo, S\~{a}o Paulo, Brazil}

\author{E. Solano}
\affiliation{Department of Physical Chemistry, University of the Basque Country UPV/EHU, Apartado 644, 48080 Bilbao, Spain}
\affiliation{IKERBASQUE, Basque Foundation for Science, Maria Diaz de Haro 3, 48013 Bilbao, Spain}
\affiliation{International Center of Quantum Artificial Intelligence for Science and Technology (QuArtist) and Department of Physics, Shanghai University, 200444 Shanghai, China}
\affiliation{IQM, Munich, Germany}

\author{M. Sanz}
\affiliation{Department of Physical Chemistry, University of the Basque Country UPV/EHU, Apartado 644, 48080 Bilbao, Spain}

\author{L. C. C\'{e}leri}
\affiliation{Institute of Physics, Federal University of Goi\'{a}s, POBOX 131, 74001-970, Goi\^{a}nia, Brazil}
\affiliation{Department of Physical Chemistry, University of the Basque Country UPV/EHU, Apartado 644, 48080 Bilbao, Spain}

\begin{abstract}
The quench dynamics of many-body quantum systems may exhibit non-analyticities in the Loschmidt echo, a phenomenon known as dynamical phase transition (DPT). Despite considerable research into the underlying mechanisms behind this phenomenon, several open questions still remain. Motivated by this, we put forth a detailed study of DPTs from the perspective of quantum phase space and entropy production, a key concept in thermodynamics. We focus on the Lipkin-Meshkov-Glick model and use spin coherent states to construct the corresponding Husimi-$Q$ quasi-probability distribution. The entropy of the $Q$-function, known as Wehrl entropy, provides a measure of the coarse-grained dynamics of the system and, therefore, evolves non-trivially even for closed systems. We show that critical quenches lead to a quasi-monotonic growth of the Wehrl entropy in time, combined with small oscillations. The former reflects the information scrambling characteristic of these transitions and serves as a measure of entropy production. On the other hand, the small oscillations imply negative entropy production rates and, therefore, signal the recurrences of the Loschmidt echo. Finally, we also study a Gaussification of the model based on a modified Holstein-Primakoff approximation. This allows us to identify the relative contribution of the low energy sector to the emergence of DPTs. The results presented in this article are relevant not only from the dynamical quantum phase transition perspective, but also for the field of quantum thermodynamics, since they point out that the Wehrl entropy can be used as a viable measure of entropy production.
\end{abstract}

\maketitle

%
%
\section{Introduction}

The dynamics of closed quantum many-body systems has been the subject of considerable interest in the last decade. After a sudden quench, for instance, the support of a local operator will in general spread through all of Hilbert space. The precise way through which this takes place reveals important information about the basic mechanisms underlying the dynamics. A particularly interesting example is the so-called dynamical phase transition (DPT), first discovered in Ref.~\cite{Heyl2013}, and subsequently explored in distinct situations. In Ref.~\cite{Vajna2015} it was shown that DPTs have no connection with standard  equilibrium phase transitions. The case of long-range interacting system was considered in Ref.~\cite{Stauber2017}, while Ref.~\cite{Schuricht2013} treated the case of nonintegrable systems. An attempt to classify dynamical phase transitions (DPTs), leading to the definition of first-order DPT, was introduced in Ref.~\cite{Eckstein2014}. Finally, we can mention a very interesting connection between these DPTs and  quantum speed limits, developed in Ref.~\cite{Heyl2017}. On the experimental side, we can mention the observation of DPTs in the Ising model in an optical lattice~\cite{Jurcevic2017}, in the topological phase in the Haldane model using an ion trap platform~\cite{Flaschner2018} and in a simulation of the Ising model in a quantum computer~\cite{Zhang2017}. We refer the reader to the recent reviews~\cite{Heyl2018,Heyl2019,Zvyagina2016} for further details about this field.

The central quantity in the theory of DPTs, which usually occur in a quenched quantum system, is the Loschmidt echo. The basic scenario consists of initially preparing a system in the ground-state $|\psi_0\rangle$ of some Hamiltonian $H_0$. At $t=0$ the system is then quenched to evolve according to a different Hamiltonian $H$. The Loschmidt echo is  defined as 
\begin{equation}
\mathcal{L}(t) = |\langle \psi_0 | \psi_t \rangle|^2 
= |\langle \psi_0 | e^{-i H t} |\psi_0 \rangle|^2, 
\label{loschmidt}
\end{equation}
and therefore, quantifies the overlap between the initial state and the evolved state at any given time. In other words, it measures how the support of the wavefunction spreads through the many-body Hilbert space.

In quantum critical systems, the Loschmidt echo in Eq.~(\ref{loschmidt}) is characterized by sharp recurrences (see Fig.~\ref{fig:rate_function}(a) for an example). The nature of these recurrences is more clearly expressed in terms of the  rate function, 
\begin{equation}
r(t) = - \frac{1}{N}\log \mathcal{L}(t), 
\label{rate}
\end{equation}
where $N$ is the system size. In the thermodynamic limit ($N\to\infty$), the rate $r(t)$ presents non-analyticities (kinks) at certain instants of time (c.f. Fig.~\ref{fig:rate_function}(b)), which are the hallmark of DPTs. The Loschmidt echo in Eq.~(\ref{loschmidt}) shows a formal relation with a thermal partition function at imaginary time, which allows one to link these non-analyticities to the Lee-Yang/Fisher~\cite{Yang1952,Fisher1965} zeros of $\mathcal{L}(t)$ (see  Ref.~\cite{Heyl2018} for more details). 

Despite considerable progress in our understanding of DPTs, several open questions still remain, for instance, regarding their universality and if they can be captured from macroscopic properties~\cite{Heyl2018,Heyl2019,Zvyagina2016}. In this article we focus on two deeply related questions. The first one concerns the transition from quantum to classical, i.e. which aspects of the problem are genuinely coherent and which could be understood from purely classical equations of motion. Defining this transition is not trivial for the great majority of models, such as spin chains. The second issue concerns which sectors of the Hamiltonian contribute to the transition. Indeed, quantum phase transitions depend only on the low energy sector (ground-state plus the first few excited states). DPTs, on the other hand, should in principle depend on the entire spectrum. 

The interpretation of $\mathcal{L}(t)$ as a measure of how the support of the system spreads in time brings a clear thermodynamic flavor to DPTs. In the language of classical thermodynamics, an expanding gas fills all of available space, causing the entropy to grow monotonically in time. In closed quantum systems, however, the von Neumann entropy is a constant of motion, despite possible information scrambling. This idea has been explored with certain detail in the context of fluctuation theorems and non-equilibrium lag~\cite{Campisi2017}. 

One way to reconcile these two views is to move to quantum phase space. In Ref.~\cite{Altland2012}, the authors studied the dynamics of the Dicke model in terms of the Husimi-$Q$ quasiprobability function. They showed that the closed (unitary) evolution yields,  notwithstanding, a diffusive-type Fokker-Planck equation in phase-space. It is, of course, a special type of diffusion in order to comply with the fact that the system is closed, and thus energy must be conserved. The Husimi function can be viewed as a convolution of the system's state with a heterodyne measurement. Consequently, it provides a coarse-grained description of the dynamics, and thus, naturally accounts for the scrambling of information.

Motivated by this, we put forth in this article a detailed study of DPTs from the optics of quantum phase space. We focus on the Lipkin-Meshkov-Glick model, describing the dynamics of a single macrospin~\cite{Lipkin1965,Meshkov1965,Glick1965,Cirac1998,Garanin1998,Latorre2005,Vidal2004,Vidal2004b,Ribeiro2007,Ribeiro2008,Ribeiro2009,Dusuel2004,Dusuel2005,Das2006,Hamdouni2007,Bao2020} (Sec.~\ref{sec:LMG_DQPT}). This model allows for a well-defined classical limit, which takes place when the spin $j \to \infty$. In this case, the model reduces to the classical dynamics of a spinning top~\cite{Ribeiro2008}. In addition, it allows for a neat construction in terms of quantum phase space by using the idea of spin coherent states: the corresponding Husimi function describes a quasiprobability distribution in the unit sphere. 

The entropy associated with the $Q$-function is known as Wehrl's entropy~\cite{Wehrl1978,Wehrl1979} and it can be attributed an operational meaning in terms of sampling through heterodyne measurements~\cite{Buzek1995}. For this reason, it upper bounds the von Neumann entropy, since it encompasses also the extra Heisenberg uncertainty related to spin coherent states. Moreover, for the same reason, the Wehrl entropy also evolves non-trivially even in closed system, unlike the von Neumann entropy, which is constant. It consequently captures the scrambling of information, very much like the Loschmidt echo in Eq.~(\ref{loschmidt}), but from the phase-space perspective. 

As we show in Sec.~\ref{sec:entro}, the dynamics of the Wehrl entropy offers valuable insight about the nature of DPTs. After a critical quench, it grows quasi-monotonically, combined with small oscillations. The growth reflects the information scrambling characteristic of DPTs. The oscillations, on the other hand, mimic the non-analytic behavior of the Loschmidt echo and reflect information backflow to the initial Hilbert space sector. Finally, in Sec.~\ref{sec:HP}, we also carry out an analysis using a generalized Holstein-Primakoff transformation that is known to faithfully capture the entire low-energy sector in the thermodynamic limit. This allows us to address which parts of the spectrum are essential for the description of DPTs. As we show, this procedure yields accurate predictions for small quenches, but fails for the strong quenches required to observe DPTs.

%
%
\section{Dynamical phase transition in the Lipkin-Meshkov-Glick model}
\label{sec:LMG_DQPT}

Let us consider the Lipkin-Meshkov-Glick (LMG) model~\cite{Lipkin1965,Meshkov1965,Glick1965}, described by the Hamiltonian
\begin{equation}
H = -h J_z - \frac{1}{2j}\gamma_x J_{x}^{2},
\label{eq:LMGmodel}
\end{equation}
where $j$ is the total angular momentum, $h \geq 0$ is the magnetic field and $\gamma_{x}>0$ (critical field $h_c = \gamma_x$). This model can be viewed as the fully connected version of a system of $N = 2j$ spin 1/2 particles (it therefore presents mean-field exponents). 
Since it is analytically tractable, it has been the subject of several studies over the last decades~\cite{Cirac1998,Garanin1998,Latorre2005,Vidal2004,Vidal2004b,Ribeiro2007,Ribeiro2008,Ribeiro2009,Dusuel2004,Dusuel2005,Das2006,Hamdouni2007,Bao2020}. 

\subsection{Brief review of the quantum phase transition}

Before describing the dynamical phase transition, let us firstly discuss the regular quantum phase transition for this model, which occurs in the thermodynamic limit, $j \rightarrow \infty$. In order to do this, it is convenient to define the so-called spin coherent state
\begin{equation}
    \ket{\Omega} = e^{-i\phi J_z} e^{-i\theta J_y}\ket{j},
    \label{eq:SCS_def}
\end{equation}
where $\ket{j}$ is the eigenstate of $J_z$ with eigenvalue $j$ and $\theta \in [0,\pi]$ and $\phi \in [0,2\pi]$ are polar coordinates. These states represent the closest quantum analog of a classical angular momentum vector of fixed length $j$, in the sense that the expectation values of the spin operators in the state of Eq.~(\ref{eq:SCS_def}) take the form 
\[
\big(\langle J_x \rangle, \langle J_y \rangle, \langle J_z \rangle\big) = j \big(\sin\theta \cos\phi, \sin\theta\sin \phi, \cos\theta\big). 
\]
Moreover, expectation values of higher powers, such as $\langle J_x^2 \rangle$, differ from $\langle J_x \rangle^2$ only by terms which become negligible in the limit of large $j$. As a consequence, it can be shown that the leading order of the ground-state of the LMG model in the thermodynamic limit is a spin coherent state for certain values of $\theta$ and $\phi$~\cite{Ribeiro2008}. 
The energy in this limit can be computed as $E=\bra{\Omega}H\ket{\Omega}$, resulting in~\cite{Castanos2006}
\begin{equation}
    \frac{E}{j} = -h\cos\theta - \frac{\gamma_x}{2}\sin^2\theta\cos^2\phi.
     \label{eq:GSenergyLMG}
\end{equation}
The corrections to this behavior will be computed explicitly in Sec.~\ref{sec:HP}.

The ground-state energy is then found by minimizing Eq.~\eqref{eq:GSenergyLMG} over $\theta$ and $\phi$, leading to the set of equations
\begin{eqnarray}
    \sin{\theta}(h-\gamma_x \cos{\theta}\cos^2\phi) &= 0, \nonumber \\[0.2cm]
    \gamma_x \sin^2\theta \cos{\phi}\sin{\phi} &= 0 \nonumber.
\end{eqnarray}
For $h > \gamma_x$ the only solution is $\theta = 0$, in which case $\phi$ is arbitrary. For $h < \gamma_x$, however, two new solutions appear, corresponding to
\begin{equation}\label{eq:theta_choice}
    \cos{\theta} = \frac{h}{\gamma_x},
\end{equation}
and $\phi = 0$ or $\phi = \pi$. The magnetization $m=\sin\theta \cos\phi$ therefore serves as the order parameter of the model. This is identically zero for $h > \gamma_x$ and $m=\sqrt{h_c^2 - h^2}/\gamma_x$ otherwise. The emergence of these new solutions identifies the critical field $h_{c} = \gamma_{x}$.

\subsection{Dynamical phase transition} 

Let us now consider the introduction of quenches in the field $h$. The system is  prepared in the ground state $\vert\psi_{0}\rangle$ of  $H_{0} = H(h_0)$, and at $t=0$, it evolved under the final  Hamiltonian $H = H(h)$. 
To quantify the DPT, we use the Loschmidt echo defined in Eq.~(\ref{loschmidt}) and the corresponding rate in Eq.~(\ref{rate}), with $N = 2j$. In this model, a subtlety arises because the ground-state is two-fold degenerate. As discussed in Appendix~\ref{app:num}, however, this introduces effects which become negligible in the thermodynamic limit. For this reason, we henceforth focus only on the analysis starting from one of the ground-states. 

\begin{center}
    \begin{figure}
        \centering
        \includegraphics[width=0.5\textwidth]{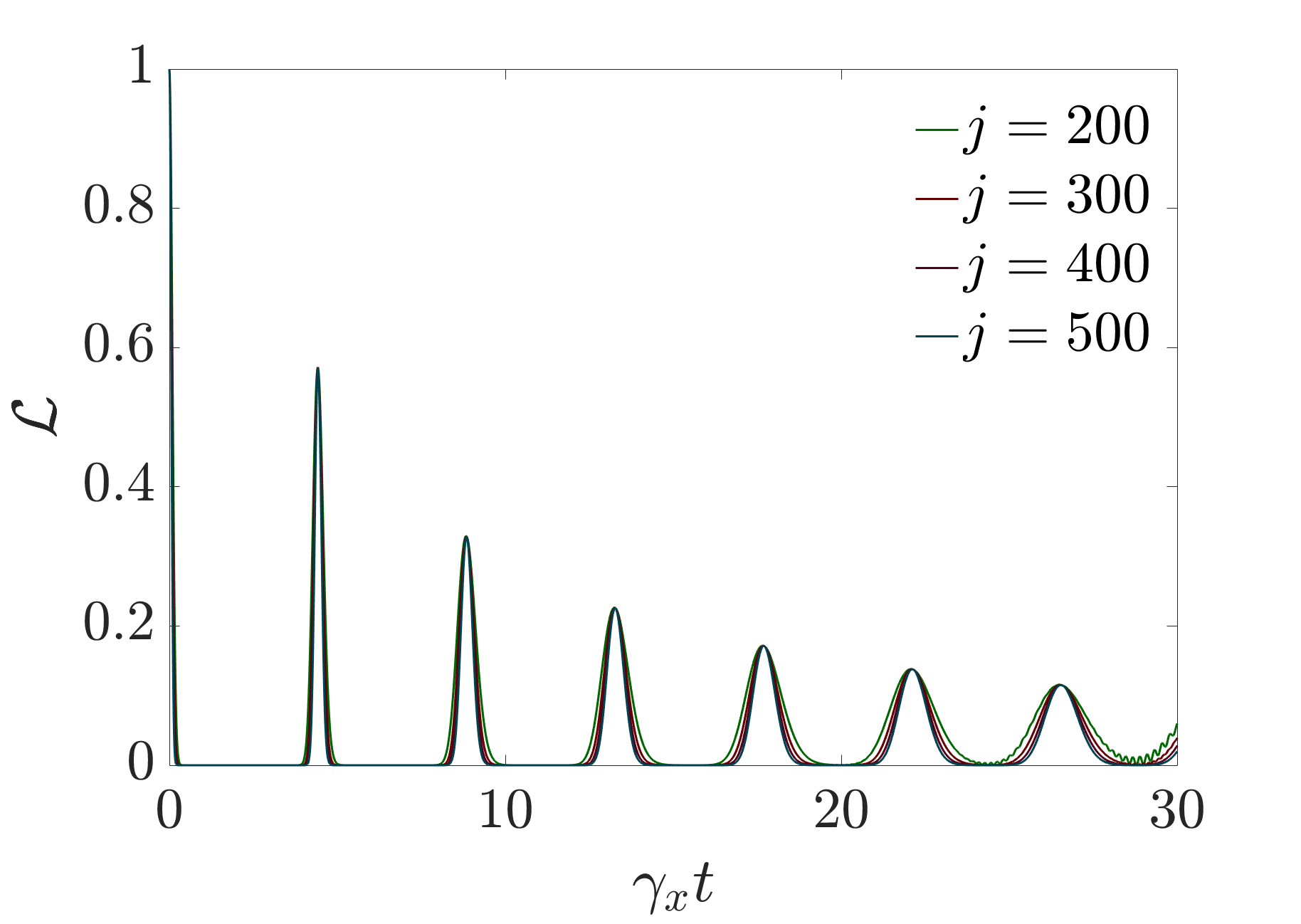} \\
        \includegraphics[width=0.5\textwidth]{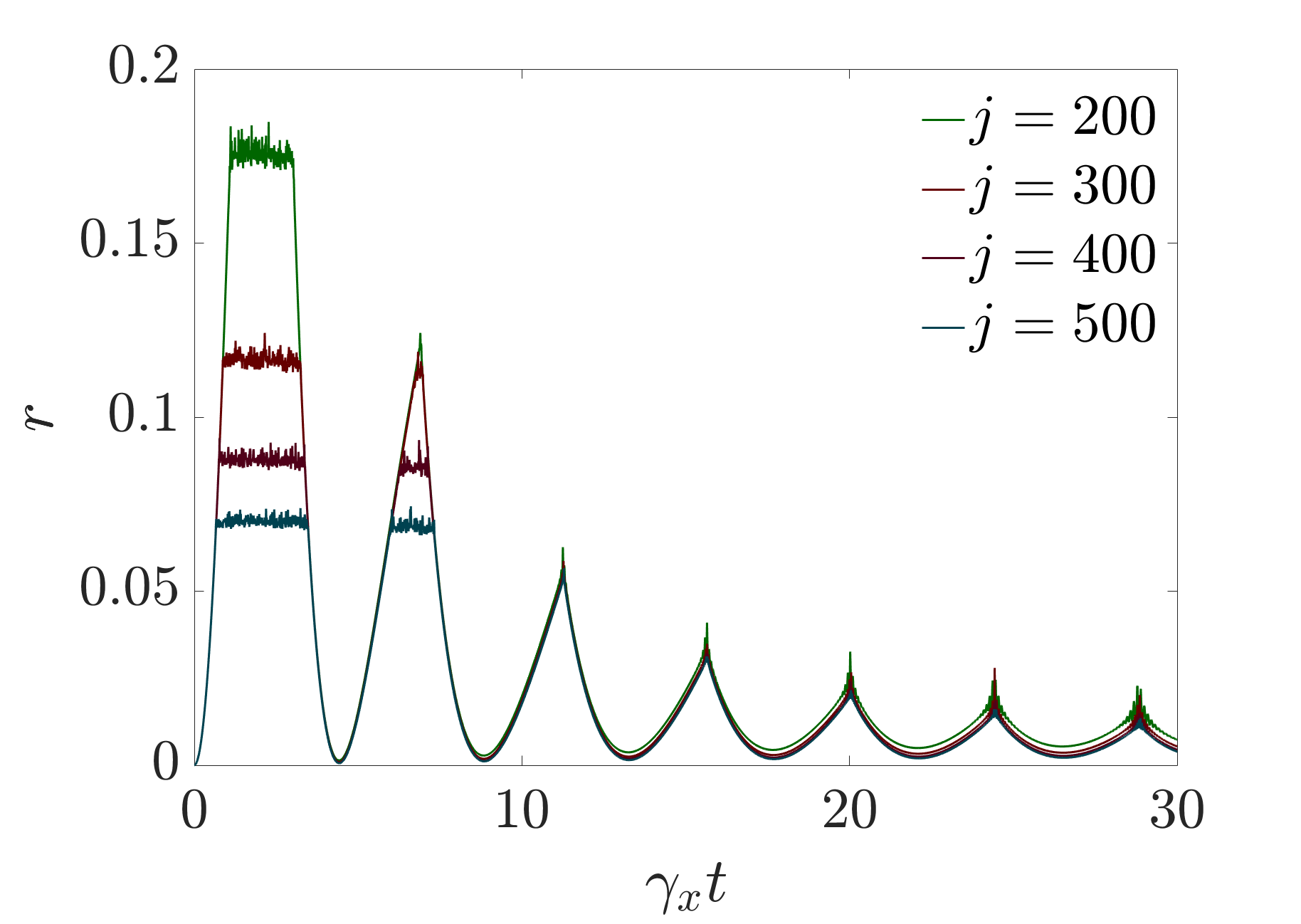}
        \caption{{Dynamical phase transition}. The top panel shows the Loschmidt echo~(\ref{loschmidt}) while the bottom one displays the dynamical behaviour of the rate function~(\ref{rate}) for a quench in the LMG model, from $h_0 = 0$ to $h = 0.8$. The curves correspond to different values of $j$. The plateaus that occur in $r(t)$ at short times is a numerical artifact. They happen because $\mathcal{L}(t)$ becomes exponentially small in these regions (even though this could be fixed with increasing the precision, this is not viable in practice because of the exponential dependence).}
        \label{fig:rate_function}
    \end{figure}
\end{center}

For the sake of concreteness, we focus on quenches from $h_0=0$ to $h< \gamma_x$. Results for the Loschmidt echo and the rate function are shown in Fig.~\ref{fig:rate_function} for several values of $j$. The echo (top panel in Fig.~\ref{fig:rate_function}) vanishes for certain periods of time, but  presents sharp periodic revivals at certain instants. This is convoluted with a damping, causing the time decay of the magnitude of $\mathcal{L}(t)$. The presence of a DPT becomes visible in the rate function (bottom panel in Fig.~\ref{fig:rate_function}), which presents kinks at certain instants of time, called critical times $t_{c}$.

%
%
\section{\label{sec:entro}Entropic dynamics in quantum phase space}

Let us now address the core part of our article, in which we put forth an analysis of the DPT in Fig.~\ref{fig:rate_function} from the perspective of quantum phase space. The spin Husimi-$Q$ function associated with an arbitrary density matrix $\rho$ is defined as
\begin{equation}
    Q(\Omega) = \bra{\Omega}\rho\ket{\Omega},
    \label{eq:Husimi_Q}
\end{equation}
where $\ket{\Omega}$ are the spin coherent states given in Eq.~(\ref{eq:SCS_def}). This quantity is always non-negative and normalized to unity according to $ (2j+1)/4\pi \int d\Omega Q(\Omega) = 1$, where $\dd\Omega = \sin\theta \dd\theta \dd\phi$. 
It therefore represents a quasi-probability distribution in the unit sphere, offering the perfect  platform to understand the quantum to classical transition~\cite{Takahashi1985}. 
In our case, $\rho = |\psi_t\rangle\langle \psi_t |$ so the Husimi function simplifies to $Q = |\langle \Omega | \psi_t \rangle|^2$. 

The entropy associated with $Q(\Omega)$ is known as Wehrl's entropy~\cite{Wehrl1978,Wehrl1979}, defined as 
\begin{equation}
    S_{Q} = -\frac{2j+ 1}{4\pi}\int \dd\Omega\; Q(\Omega)\ln{Q(\Omega)}.
    \label{eq:wehrl}
\end{equation}
An operational interpretation of this quantity in terms of sampling through heterodyne measurements was given in Ref.~\cite{Buzek1995}. It has also been applied in different contexts, like entanglement theory~\cite{Mintert2004}, uncertainty relations~\cite{Palma2018}, and quantum phase transitions~\cite{Castanos2015}, just to name a few. In the context of thermodynamics, our interest here, a theory of entropy production for spin systems was put forth in Ref.~\cite{Santos2018} and the corresponding bosonic analog in Ref.~\cite{Goes2020}.

In classical thermodynamics, the entropy of a \emph{closed} system should be monotonically increasing with time. The quantity
\begin{equation}
\dv{S_{Q}}{t}:= \Pi_{Q}
\label{eq:entro_prod}
\end{equation}
is interpreted as the entropy production rate, since it reflects the entropy that is irreversibly produced in the system. The second law then states that $\Pi_Q \geq 0$. In open systems, on the other hand, one  has instead 
\begin{equation}
    \dv{S_{Q}}{t} = \Pi_{Q} - \Phi_Q,
\end{equation}
where $\Phi_Q$ is the entropy flux rate from the system to the environment. In open systems, $\dd S_Q/\dd t$ does not have a well-defined sign, since the flux $\Phi_Q$ can be arbitrary. However, one still has $\Pi_Q \geq 0$. In the present case there is no associated flux ($\Phi_Q = 0$), since the dynamics of the system is closed. Moreover, it is not guaranteed that $\dd S_Q/\dd t \geq 0$ for all times, since our system is not in the thermodynamic limit. Notwithstanding, one still expects that strong information scrambling, as happens in critical systems, should lead to a $\Pi_Q$ which is \emph{most of the time} non-negative. 

\begin{center}
    \begin{figure}
        \centering
        \includegraphics[width=0.5\textwidth]{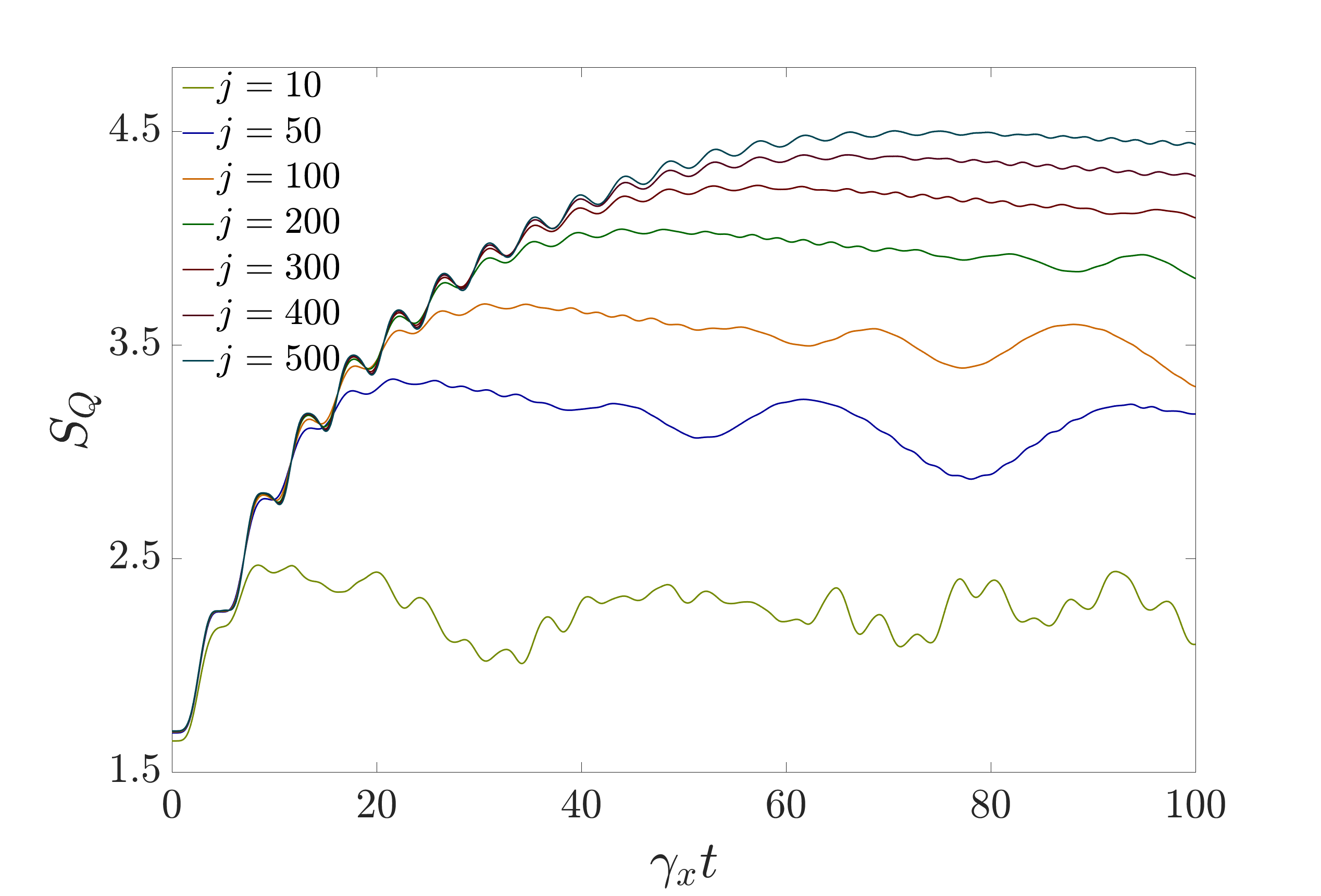} \\
        \includegraphics[width=0.5\textwidth]{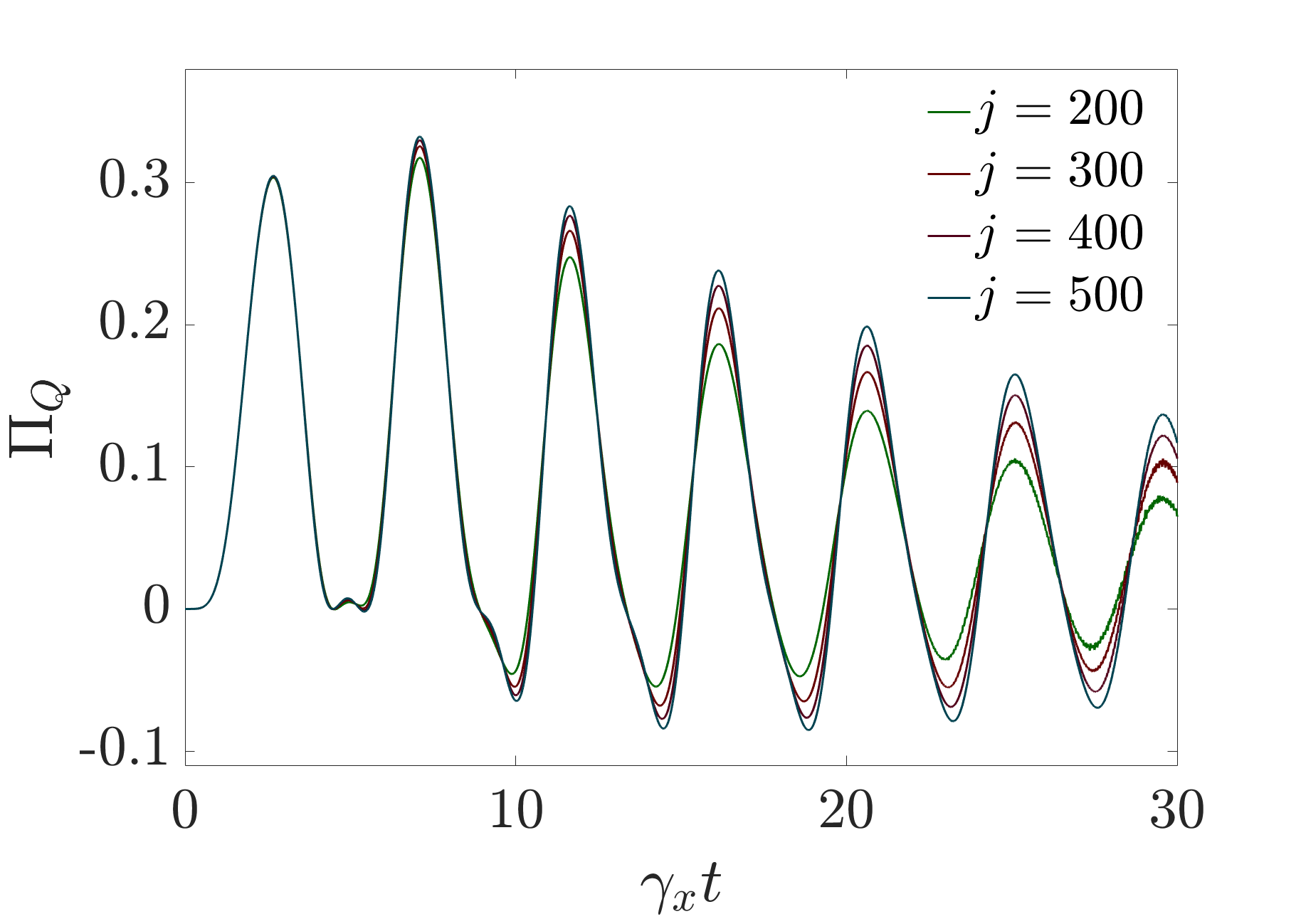}
        \caption{\textbf{Entropy dynamics}. The top panel shows the dynamics of the Wehrl entropy~(\ref{eq:wehrl}) while the entropy production rate, \eqref{eq:entro_prod}, is displayed in the bottom panel, for the same quench protocol used in Fig.~\ref{fig:rate_function}. }
        \label{fig:entropy_j}
    \end{figure}
\end{center}

Numerical results for the Wehrl entropy in Eq.~(\ref{eq:wehrl}) and the entropy production rate in Eq.~(\ref{eq:entro_prod}) are shown in Fig.~\ref{fig:entropy_j} for the same quench protocol used in Fig.~\ref{fig:rate_function}, which are remarkable. As $j$ increases, the Wehrl entropy presents small oscillations enveloping a monotonically increasing behavior. This increase clearly reflects the information scrambling characteristic of the DPT. It shows that, as time passes, the coarse-grained nature of $Q(\Omega)$ causes the available information about the system's state to be degraded as a function of time. For long times and $j$ sufficiently large, $S_Q$ has a tendency to saturate at a constant value. As anticipated, $\Pi_Q$ oscillates in time, being predominantly positive, but also becoming negative at certain times. These negativities represent the backflow of information, which is characteristic of the recurrences in $\mathcal{L}(t)$ (see Fig.~\ref{fig:rate_function}).

\begin{center}
    \begin{figure}
        \centering
        \includegraphics[width=0.5\textwidth]{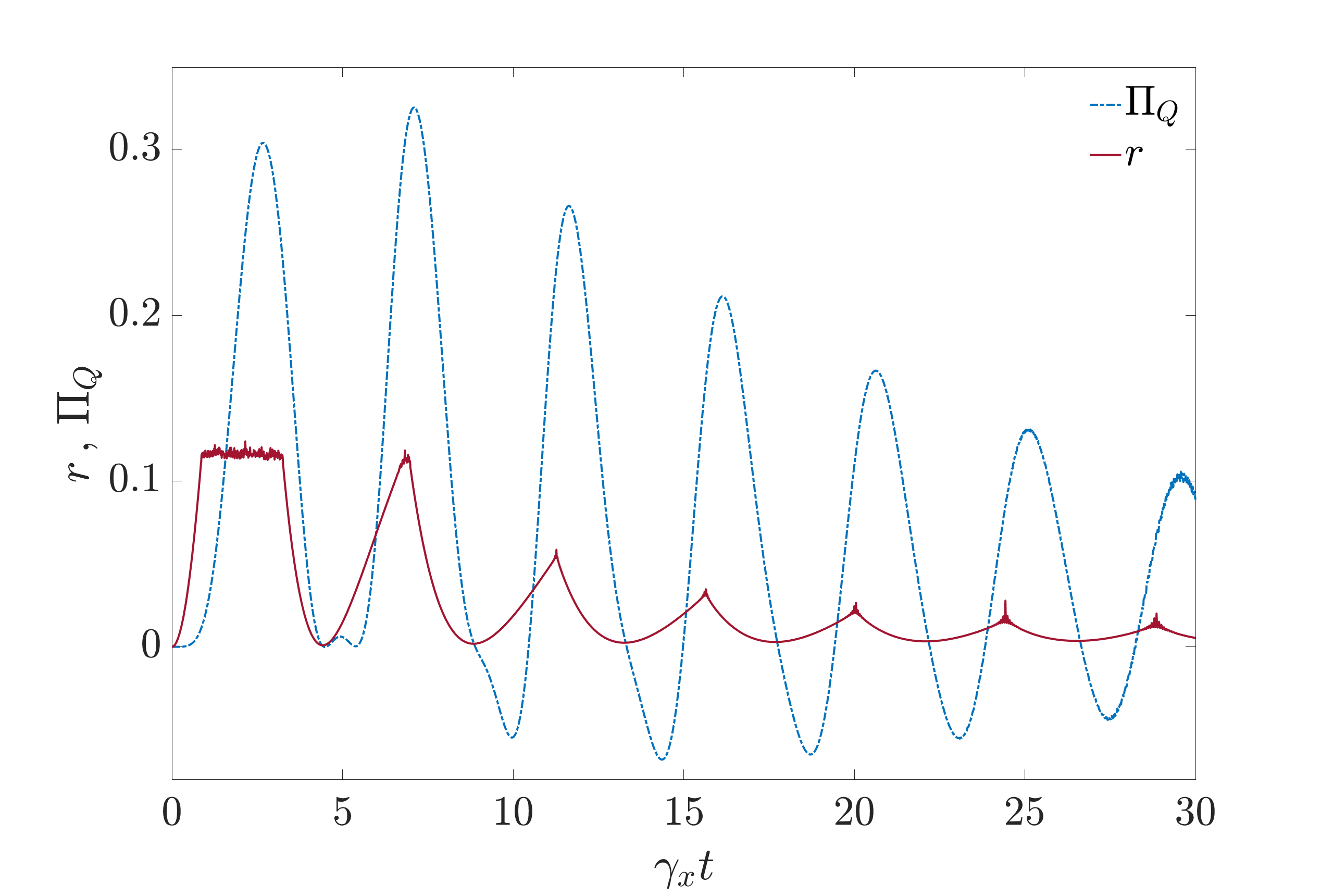}
        \caption{Rate function $r(t)$, along with the entropy production rate $\Pi_{Q}$, for  $j=300$.
        Other parameters are the same as in Fig.~\ref{fig:rate_function}.
        }
        \label{fig:rate_entro}
    \end{figure}
\end{center}

\begin{center}
    \begin{figure}
        \centering
        \includegraphics[width=0.5\textwidth]{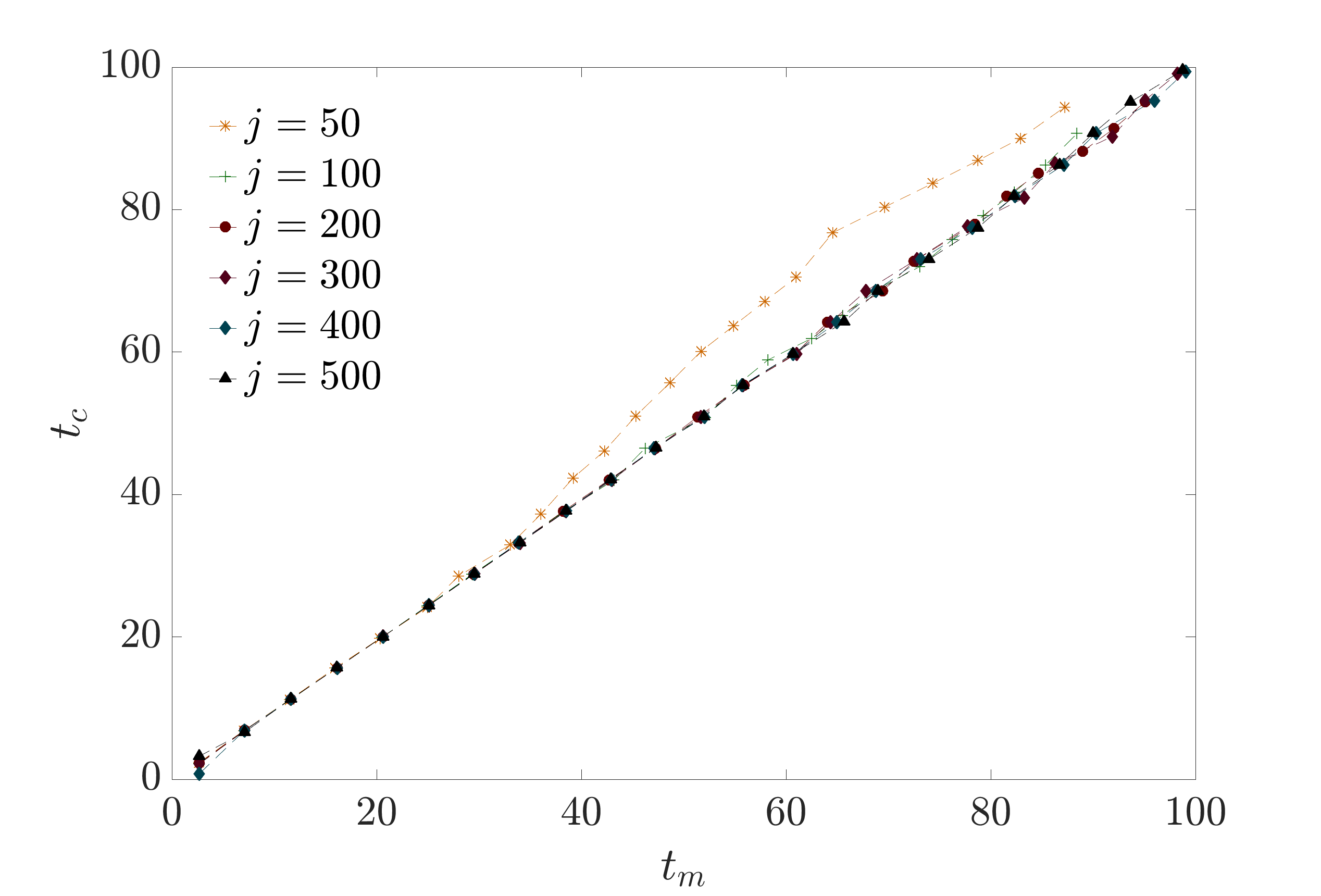}
        \caption{Critical times $t_{c}$, associated with the non-analytical behaviour of $r(t)$ and the times $t_{m}$ corresponding to local maxima of $\Pi_Q$.
        Other parameters are the same as in Fig.~\ref{fig:rate_function}.
        }
        \label{fig:critical}
    \end{figure}
\end{center}

Figure~\ref{fig:rate_entro} shows the entropy production rate $\Pi_Q$ along with the rate function $r(t)$ for $j = 300$. As we can see, their behaviours are clearly linked. In order to make such relation clearer, we present in Fig. \ref{fig:critical} the relation between the critical times $t_{c}$ where the dynamical quantum phase transitions occur, i.e. the non-analytical points of $r(t)$, and the times $t_{m}$ at which $\Pi_Q$ present local maxima. From this result it becomes clear that, for large $j$, one approaches $t_{c} \approx  t_{m}$, showing how the maxima of $\Pi_Q$ perfectly correlate with the critical times. This corroborates the idea that the oscillations in $\Pi_Q$ indeed reflect the DPT. A more detailed analysis is presented in Appendix \ref{app:num}.

Finding a closed-form expression for the entropy production rate~(\ref{eq:entro_prod}) is in general not possible. Notwithstanding, we were able to identify which parts of the dynamics, in terms of the associated Fokker-Planck equation, contribute to $\Pi_Q$. This analysis, however, is cumbersome and it is thus postponed to Appendix~\ref{app:entro}. 

%
%
\section{\label{sec:HP}Contribution from the low energy sector}

\subsection{Holstein-Primakoff approximation for the LMG model} 

Equilibrium quantum phase transitions are almost entirely described by the low energy sector, i.e. ground state plus the first few excited states. This is not true for DPTs which, in principle, depend on the entire spectrum. In this section we ask which aspects of DPTs can nonetheless be captured by the low energy sector. We do this by introducing a Gaussification procedure based on a generalized Holstein-Primakoff (HP) representation \cite{Holstein1940}. Finally, we then compare the predictions of this HP model with the full numerics studied in the previous section. 

The HP method represents spin operators in terms of a single bosonic mode, by means of the non-linear transformation~\cite{Holstein1940}
\begin{subequations}
    \begin{equation}
         J_z = j -\dg{a}a,
         \label{eq:HP_JZ}
    \end{equation}
    \begin{equation}
        J_+ = \sqrt{2j-\dg{a}a}\;a,
        \label{eq:HP_Jp}
    \end{equation}
\end{subequations}
where $a$ is a bosonic operator satisfying $[a,a^\dagger] = 1$. Equation~(\ref{eq:HP_JZ}) shows that the excitations of $a^\dagger a$ are mapped onto excitations of $J_z$, starting from the state $|j\rangle$, downwards. This is reasonable when $h > h_c$. But when $h < h_c$, the ground-state will not be close to $|j\rangle$ at all. To take this into account, we first rotate the Hamiltonian~(\ref{eq:LMGmodel}) by an angle $\theta$  around the $y$-axis, before applying the HP transformation. That is, we first consider the rotated Hamiltonian 
\begin{IEEEeqnarray}{rCl}
 \label{eq:RotatedHamiltonian}
H_R &=& e^{i \theta S_y} H e^{-i \theta S_y} \\[0.2cm]
&=&-h(J_z\cos\theta - J_x\sin\theta) -\frac{\gamma_x}{2j}(J_z\sin\theta + J_x\cos\theta)^2,
   \nonumber
\end{IEEEeqnarray}
where the value of $\theta$ will be fixed below. We now introduce the HP transformation on $H_R$ instead of $H$. Expanding for large $j$ and keeping only terms which are at most quadratic in $a$ and $a^\dagger$, one finds
\begin{IEEEeqnarray}{rCl}
\label{HP_approx_1}
H_R &=& E - \frac{\sqrt{2j}}{2} \sin\theta (\gamma_x\cos\theta - h) (a+a^\dagger) \\[0.2cm]
\nonumber
&&+ a^\dagger a (\gamma_x \sin^2 \theta + h \cos\theta) - \frac{\gamma_x}{4} \cos^2\theta (a+a^\dagger)^2,
\end{IEEEeqnarray}
where $E$ is the classical energy given in Eq.~(\ref{eq:GSenergyLMG}) with $\phi = 0$. The role of $\phi$ is trivial, it is not necessary to rotate around the $z$ axis as well. 

We can now  choose $\theta$ to eliminate the linear term  proportional to $a+a^\dagger$. This leads to the same choice that minimized the classical energy in Eq.~(\ref{eq:theta_choice}), i.e. 
\begin{equation}\label{HP_theta_values}
    \theta_h = \begin{cases}
    \arccos(h/\gamma_x), & \text{ if } h < h_c = \gamma_x \\[0.2cm]
    0 & \text{ otherwise}.
    \end{cases}
\end{equation}
With this choice, Eq.~(\ref{HP_approx_1}) reduces to 
\begin{equation}\label{HR_below}
    H_R(h) = E + \gamma_x a^\dagger a - \frac{h^2}{4\gamma_x} (a+a^\dagger)^2, \qquad h < h_c,
\end{equation}
and 
\begin{equation}\label{HR_above}
    H_R(h) = E + h a^\dagger a - \frac{\gamma_x}{4} (a+a)^\dagger,\qquad h > h_c.
\end{equation}
Therorefor, the HP method leds to the same classical energy landscape as using spin coherent states [Eq.~(\ref{eq:GSenergyLMG})]. However, it is important to remark that it yields an operator-based representation of the fluctuations around the ground-state. Note how $E \sim j$ is extensive, whereas the fluctuations in Eqs.~(\ref{HR_below}) and (\ref{HR_above}) are independent of $j$. Notwithstanding, as we will find below, this does not mean the fluctuations are negligible when they become significantly close to criticality. 

We now use the above results to determine the ground-state and the energy gap between the ground-state and the first excited state. To do this, we introduce the squeeze operator $S_{\xi} = e^{\frac{\xi}{2}(a^2 - (a^\dagger)^2)}$, where for our purposes it suffices to take $\xi>0$. 
This allows us to write Eqs.~(\ref{HR_below}) and (\ref{HR_above}), up to constants, as
\begin{equation}\label{HP_HR_diag}
    H_R(h) = E + \omega_h S_{h}^\dagger a^\dagger a S_{h},
\end{equation}
where 
\begin{equation}\label{HP_omega_values}
    \omega_h = \begin{cases}
    \sqrt{\gamma_x^2 - h^2}, & \text{ if } h < h_c, \\[0.3cm]
    \sqrt{h (h- \gamma_x)}, & \text{ otherwise}, 
    \end{cases}
\end{equation}
and $S_h = S(\xi_h)$ with 
\begin{equation}\label{HP_rh_values}
    \xi_h = \begin{cases}
    - \frac{1}{4} \ln (1-h^2/\gamma_x^2), & \text{ if } h < h_c, \\[0.3cm]
    - \frac{1}{4} \ln (1 - \gamma_x/h), & \text{ otherwise}. 
    \end{cases}
\end{equation}
The Hamiltonian~(\ref{HP_HR_diag}) is now diagonal, so that $\omega_h$ describes precisely the energy level spacing of the first few excited levels. Whence, we conclude that, at low energies, the excitations are equally spaced with energy gap $\omega_h$. As a feature of quantum phase transitions, the gap closes at $h = h_c$. Moreover, it does so from both directions and in a manner which is not symmetric in $h$. 

Finally, to compute the ground-state, we must first go back to the original Hamiltonian by undoing the rotation in Eq.~(\ref{eq:RotatedHamiltonian}). In the language of the HP transformation~(\ref{eq:HP_Jp}), a rotation around $y$ becomes a displacement of the bosonic mode $a$
\begin{equation}
    e^{i \theta J_y} = D(\alpha_h) =  e^{\alpha_h a^\dagger - \alpha_h^* a},
\end{equation}
where $\alpha_h = -\sqrt{2j}\; \theta_h/2$ and $\theta_h$ is given in Eq.~(\ref{HP_theta_values}). By combining Eqs.~(\ref{eq:RotatedHamiltonian}) and~(\ref{HP_HR_diag}), then it yields the original Hamiltonian in the form 
\begin{equation}\label{HP_H_final}
    H =  D_h^\dagger S_h^\dagger \big[\omega_h a^\dagger a \big] S_h D_h,
\end{equation}
where $D_h = D(\alpha_h)$.
The ground-state is now readily found to be 
\begin{equation}\label{HP_gs}
    |\psi_\text{gs}(h)\rangle = D_h^\dagger S_h^\dagger |0\rangle, 
\end{equation}
where $|0\rangle$ is the vacuum defined by $a$. The ground-state is consequently a displaced squeezed state. The amount of displacement, $\alpha_h = - \sqrt{2j} \theta_h/2$ is zero when $h>h_c$ and non-zero otherwise. Moreover, the displacement scales as $\sqrt{j}$ and thus become extremely large in the thermodynamic limit. This displacement is precisely the prediction using spin coherent states, but written in a bosonic language. In addition, the state is also squeezed by the amount $\xi_h$ given in Eq.~(\ref{HP_rh_values}). This therefore represents a correction on top of the spin coherent state predictions. The squeezing is $j$ independent, but diverges at criticality. Thus, very close to the critical point, the state should differ significantly from a spin coherent state.  

\subsection{Loschmidt echo}

Armed with the aforementined results, we can now readily compute both the Loschmidt echo and the rate function within the HP approximation. The system is initially prepared in the ground-state $|\psi_\text{gs}(h_0)\rangle$ corresponding to a field $h_0$. After the quench, the field changes to $h$. Since $S_h$ and $D_h$ are unitary, the time-evolution operator can be written as 
\begin{equation}
    e^{-i H t} = D_h^\dagger S_h^\dagger e^{-i \omega_h t a^\dagger a} S_h D_h. 
\end{equation}
The evolved state at time $t$ will then be 
\begin{equation}
    |\psi_t \rangle = e^{-i H t} | \psi_0 \rangle = D_h^\dagger S_h^\dagger e^{-i \omega_h t a^\dagger a} S_h D_h D_{h_0}^\dagger S_{h_0}^\dagger |0\rangle.  
\end{equation}
The Loschmidt echo $\mathcal{L}(t) = |\langle \psi_0 | \psi_t\rangle|^2$ becomes
\[
\mathcal{L}(t) = |\langle 0| S_{h_0} D_{h_0} D_h^\dagger \dg{S}_h e^{-i \omega_h t a^\dagger a} S_h D_h D_{h_0}^\dagger S_{h_0}^\dagger |0\rangle|^2.
\]
This can be simplified by exploiting the algebra of displacement and squeeze operators, which in this case is facilitated by the fact that the arguments $\alpha_h$ and $\xi_h$ are all real. First, one has  $D_h D_{h_0}^\dagger = D(\delta \alpha)$, where $\delta \alpha = - \sqrt{2j} (\theta_h - \theta_{h_0})/2$. Second, $D(\delta \alpha) S_{h_0}^\dagger = S_{h_0}^\dagger D(\delta\tilde{\alpha})$, where $\delta\tilde{\alpha} = \delta \alpha  e^{-\xi_{h_0}}$. Finally, we combine the two squeezing operations as $S_h S_{h_0}^\dagger = S(\delta \xi)$, where $\delta \xi = \xi_h - \xi_{h_0}$. This leads to 
\begin{equation}
    \mathcal{L}(t) = | \langle 0 | D^\dagger( \delta \tilde{\alpha} ) S^\dagger(\delta \xi) e^{-i \omega_h t a^\dagger a} S(\delta \xi) D(\delta \tilde{\alpha}) |0 \rangle |^2. 
\end{equation}
This expression can now be evaluated by noting that it represents the vacuum expectation of a thermal squeezed displaced Gaussian state at imaginary temperature $\beta = -i \omega_h$. The result is therefore 
\begin{equation}\label{HP_loch_final}
    \mathcal{L}(t) = \frac{\exp\big\{-\frac{2 j (\theta_h - \theta_{h_0}) e^{-2 \xi_{h_0}}}{1 + 4 e^{\delta \xi} \cot^2(\omega_h t/2)}\big\}}{\sqrt{\cos ^2(\omega_h t)+\cosh ^2(2 \delta \xi) \sin ^2(\omega_h t)}} ,
\end{equation}
an expression which depends only on the HP parameters $\theta_h$, $\xi_h$ and $\omega_h$, given by Eqs.~(\ref{HP_theta_values}), (\ref{HP_omega_values}) and~(\ref{HP_rh_values}), respectively. 

Note also how the exponent in~(\ref{HP_loch_final}) is extensive in $j$. As a consequence, the rate function~(\ref{rate}) with $N = 2j$ becomes, in the thermodynamic limit,  
\begin{equation}\label{HP_r_final}
    r(t) = \frac{(\theta_h - \theta_{h_0}) e^{-2 \xi_{h_0}}}{1 + 4 e^{\delta \xi} \cot^2(\omega_h t/2)}.
\end{equation}

\begin{figure}
    \centering
    \includegraphics[width=0.45\textwidth]{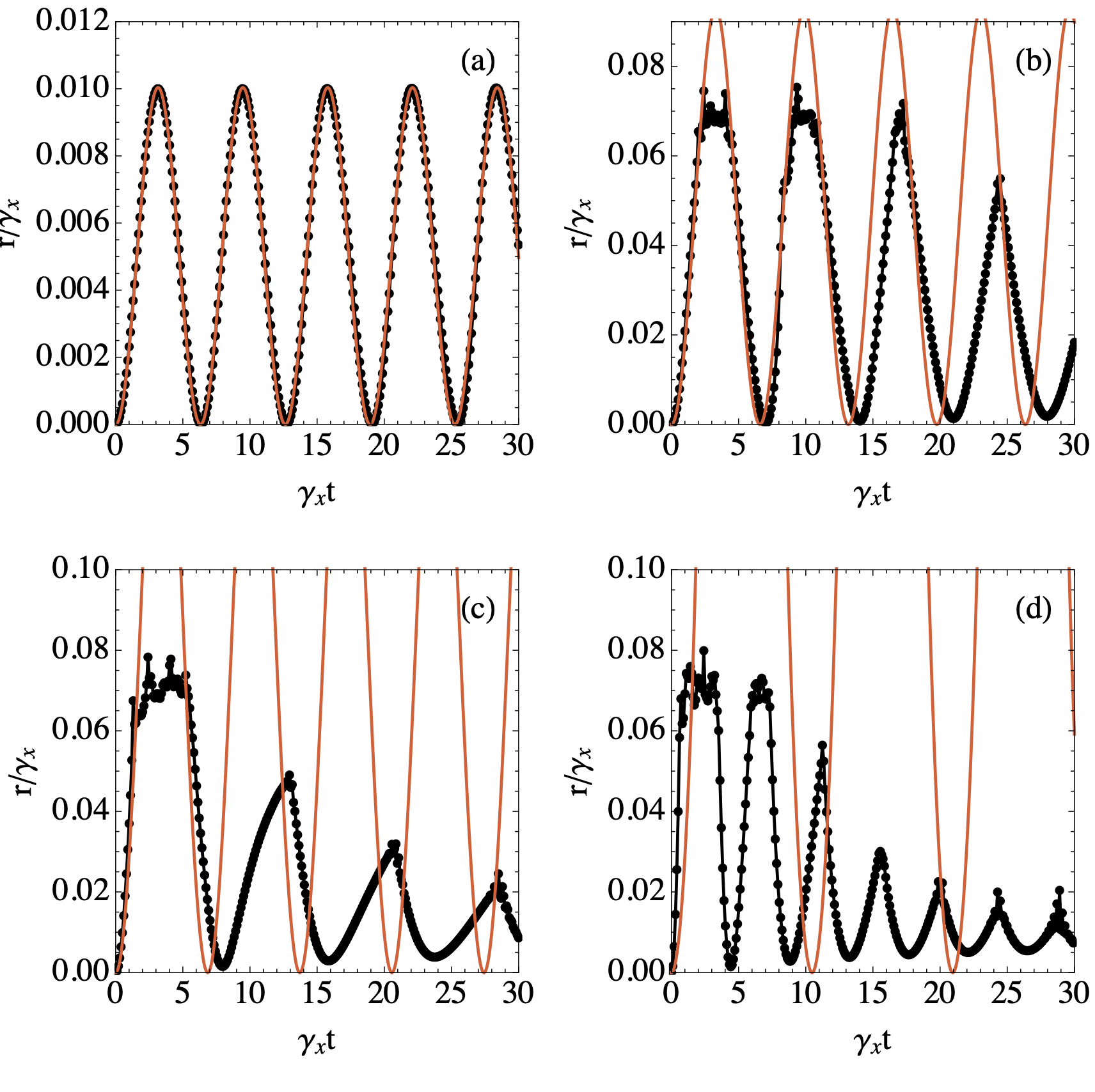}
    \caption{Comparison of the rate function $r(t)$ between the full numerics (black points) and  the Holstein-Primakoff approximation [Eq.~(\ref{HP_r_final})] for $j = 200$. Each curve describes a quench from $h_0 = 0$ to (a) $h/\gamma_x= 0.1$, (b) $h/\gamma_x = 0.3$, (c) $h/\gamma_x = 0.4$ and (d) $h/\gamma_x = 0.8$ [c.f. Fig.~\ref{fig:rate_function}(b)]. 
    }
    \label{fig:HP_loch}
\end{figure}

We can now finally address the main question posed in the beginning of this section. Namely, what is the contribution of the low energy sector to DPTs. To do this, we simply compare Eq.~(\ref{HP_r_final}) with the full numerics. The results are presented in Fig.~\ref{fig:HP_loch}. They clearly that DPTs cannot be explained making use only of the low energy sector. 
In Fig.~\ref{fig:HP_loch}(a), for instance, we compare quenches from $h_0 = 0$ to $h/\gamma_x = 0.1$. In this case, Eq.~(\ref{HP_r_final}) (in red) faithfully captures the physics of the full numerical solution (black points). However, for these small quenches, the non-analyticities of $r(t)$ are not yet present. Conversely, as the value of $h$ increases, the signature kinks of the DPTs start to appear, whereas Eq.~(\ref{HP_r_final}) remains perfectly analytical. In fact, Eq.~(\ref{HP_r_final}) can only become non-analytic when $h = \gamma_x = 1$, in which case $\omega_h \to 0$ and the cotangent diverges. At this point, the high energy sector becomes so important that, albeit non-analytic, Eq.~(\ref{HP_r_final}) cannot capture at all the physics of the problem. 

%
%

\section{Conclusion}

In summary, we studied the dynamical behaviour of the entropy production in a closed system that undergoes a dynamical quantum phase transition. Specifically, we considered sudden quenches in the Lipkin-Meshkov-Glick model under the perspective of the Husimi-$Q$ quasi-probability distribution. Such approach allowed us to define the entropy associated with the $Q$-function, known as Wehrl entropy, which is a measure of the coarse-grained dynamics of the system.

From such phase space approach, we were able to show that critical quenches lead to a quasi-monotonic growth of the Wehrl entropy in time, thus demonstrating the scrambling of quantum information, a characteristic feature of these transitions. Moreover, the entropy rate also presents small oscillations, implying negative entropy production rates at certain instants, which signals the recurrences of the Loschmidt echo. These results are relevant not only from the dynamical quantum phase transition perspective, but also for the field of quantum thermodynamics, since they point out that the Wehrl entropy can be used as a viable measure of entropy production.

Finally, based on a modified Holstein-Primakoff approximation, a type of Gaussification method, we addressed the contribution of the low energy sector to the dynamical quantum phase transition. This procedure, which is known to faithfully capture the entire low energy sector in the thermodynamic limit, fails to accurately predict the behaviour of the system under dynamical quantum phase transitions.

Despite considerable research into the underlying mechanisms behind this phenomenon, several open questions still remain and we believe that the phase-space perspective that we put forth here may contribute to deepen our understanding, specially regarding the thermodynamics of such phase transitions, as indicated by the strong connection between critical times, characterizing the critical transition, and the entropy production rate, a key concept in thermodynamics.

\vspace{0.2cm}
\noindent
{\bf Acknowledgments} ---  ES, MS and LCC acknowledge support from Spanish MCIU/AEI/FEDER (PGC2018-095113-B-I00), Basque Government IT986-16, the projects QMiCS (820505) and OpenSuperQ (820363) of the EU Flagship on Quantum Technologies and the EU FET Open Grant Quromorphic and the U.S. Department of Energy, Office of Science, Office of Advanced Scientific Computing Research (ASCR) quantum algorithm teams program, under field work proposal number ERKJ333. LCC also thanks the Brazilian Agencies CNPq, FAPEG. LCC and GTL acknowledge  the Brazilian National Institute of Science and Technology of Quantum Information (INCT/IQ) for the financial support. This study was financed in part by the Coordena\c{c}\~{a}o de Aperfeiçoamento de Pessoal de N\'{i}vel Superior - Brasil (CAPES) - Finance Code 001.
GTL acknowledges the support from the S\~ao Paulo Research Foundation FAPESP (grants 2017/07973-5, 2017/50304-7 and 2018/12813-0).

%
%
\appendix

\section{Entropy production for LMG model}
\label{app:entro}

In this appendix we analytically compute the entropy production rate for the LMG model Eq.~\eqref{eq:LMGmodel}. To accomplish this task, we use the Schwinger map to transform spin operators into bosonic operators~\cite{Takahashi1975,Santos2018}. This map employs two sets of bosonic operators, $a$ and $b$, such that
\begin{equation}\label{Schwinger_map}
\begin{split}
    J_z &= \frac{1}{2}(\dg{a}a-\dg{b}b)\\
    J_+ &= \dg{J_-} = \dg{a}b.
\end{split}
\end{equation}
To simplify notation we will employ the bosonic coherent state representation for the Husimi function. In this way, we define $\alpha$ and $\beta$ as the complex amplitudes associated with operators $a$ and $b$, respectively. The correspondence table between bosonic operator acting on a state $\rho$ and a differential operator acting on the Husimi Q-function $Q(\vb{\alpha,\beta})$ is
\begin{equation}\label{eqn:Q_func_correspondence_table} 
\begin{split}
     o_i\rho &\rightarrow (\mu_i+ \pd{\cj{\mu}_i})Q(\mu_i,\cj{\mu}_i)  \\ 
    \dg{o}_i\rho &\rightarrow \cj{\mu}_iQ(\mu_i,\cj{\mu}_i)  \\ 
    \rho o_i &\rightarrow  \mu_i Q(\mu_i,\cj{\mu}_i)  \\
    \rho\dg{o}_i &\rightarrow  (\cj{\mu}_i + \pd{\mu_i})Q(\mu_i,\cj{\mu}_i),
\end{split}
\end{equation}
where $o_i$ stands for $a$ or $b$ and $\mu_i$ represents either $\alpha$ or $\beta$, $\cj{\mu}_i$ denotes complex conjugation. Using Eqs.~\eqref{Schwinger_map} and \eqref{eqn:Q_func_correspondence_table} one finds,
\begin{widetext}
\begin{equation}
\label{eqn:Z_differential_operator}
 [J_z,\rho] \rightarrow \mathcal{J}_z= \frac{1}{2}[(\alpha\pd{\alpha}Q - \cj{\alpha}\pd{\cj{\alpha}}{Q})-(\beta\pd{\beta}Q - \cj{\beta}\pd{\cj{\beta}}{Q})],
\end{equation}
and, since $2 J_x = J_+ + J_-$, it follows that
\begin{equation}
    \label{eqn:X2_differential_operator}
    \begin{split}
        4[J_x^2,\rho] \rightarrow 
        4\mathcal{J}_{x^2}
        &= (2|\alpha|^2+1)(\cj{\beta}\pd{\cj{\beta}}{Q} -\beta\pd{\beta}{Q}) +(2|\beta|^2+1)(\cj{\alpha}\pd{\cj{\alpha}}{Q} -\alpha\pd{\alpha}{Q})\\ &
        + 2[(\cj{\alpha}^2\beta\pd{\cj{\beta}}{Q}-\alpha^2 \cj{\beta}\pd{\beta}{Q}) +(\alpha\cj{\beta}^2\pd{\cj{\alpha}}{Q}-\cj{\alpha} \beta^2\pd{\alpha}{Q})]\\
        &+2(\cj{\alpha}\cj{\beta}\pd{\cj{\alpha}}{\pd{\cj{\beta}}{Q} - \alpha \beta \pd{\alpha}{\pd{\beta}{Q}}})\\
        &+ [(\cj{\alpha}^2\dpd{\cj{\beta}}{Q}-\alpha^2\dpd{\beta}{Q}) + (\cj{\beta}^2\dpd{\cj{\alpha}}{Q}-\beta^2\dpd{\alpha}{Q})]\\
        &=4(\mathcal{J}_{x^2}^{(1)}+\mathcal{J}_{x^2}^{(2)}+\mathcal{J}_{x^2}^{(3)}+\mathcal{J}_{x^2}^{(4)}) .
    \end{split}
\end{equation}
\end{widetext}
where we introduced the notation $\mathcal{J}_{x^2}^{(i)}$ representing the $i$-th line of Eq~\eqref{eqn:X2_differential_operator}.
The dynamics of our system is governed by the von Neumann equation,
\begin{equation*}
    \pd{t}{\rho} = -i[H_{x^2}+H_z,\rho] = ih[J_z,\rho] +\frac{i\gamma_x}{2j}[J_x^2,\rho],
\end{equation*}
which is mapped into a quantum Fokker-Planck equation for the Husimi Q-function of the bosonic operators introduced by the Schwinger map~\eqref{Schwinger_map},
\begin{equation*}
    \pd{t}{Q}(\vb{\mu}) = \mathcal{U}_{x^2}+\mathcal{U}_z,
\end{equation*}
where $\mathcal{U}_z = ih \mathcal{J}_z $ and $\mathcal{U}_{x^2} = (i\gamma_x/8j)\mathcal{J}_{x^2}$. 

We are interested in the entropy rate, which in the Schwinger representation is given by
\begin{equation}
\label{AppA:entropy_rate}
    \frac{\dd S_{Q}}{\dd t} = - \int \dd^4\mu\; \mathcal{U}(Q)\ln{Q} = - (\Lambda_{x^2} + \Lambda_z)
\end{equation}
where $\dd^4\mu = \dd^2\alpha \dd^2\beta $ and $\Lambda_a = \int \dd^4\mu\; \mathcal{U}_a\ln Q$ with $a=z,x^2$. From Eqs.~\eqref{eqn:Z_differential_operator} and \eqref{eqn:X2_differential_operator} one can compute the terms of Eq.~\eqref{AppA:entropy_rate}. We note that the Schwinger transformation Eq.~\eqref{Schwinger_map} maps a bounded set of spin operators into, what is in principle, an unbounded set of bosonic operators. However,  the map introduces a restriction on the set of bosonic operators, which makes the Husimi Q-function and its derivatives vanish for $\alpha,\beta \rightarrow \pm \infty$ once they are restricted, \textit{i.e.} non-zero only to a specific region of  phase space. Using this fact along with integration by parts, we find that $\Lambda_z = 0$ and the only terms that contribute to the entropy rate are those related to diffusion, \textit{i.e.} the terms with second derivatives in $\mathcal{J}_{x^2}$. $\Lambda_{x^2}$ can be independently computed for each term appearing in Eq. \eqref{eqn:X2_differential_operator}. For the first contribution we have,
\begin{widetext}
\begin{equation}
\label{eqn:LambdaX1}
\begin{split}
    \Lambda_{x^2}^{(1)} &\propto \int \dd^4\mu\; (2|\alpha|^2+1)\ln Q(\cj{\beta}\pd{\cj{\beta}}{Q}-\beta\pd{\beta}{Q}) \\
    &= \left[(2|\alpha|^2+1)\cj{\beta}Q\ln Q  - \text{c.c.} \right]_{-\infty}^{\infty} -\int \dd^4\mu\; (2|\alpha|^2+1)\bigpar{\ln Q +\cj{\beta}\frac{\pd{\cj{\beta}}{Q}}{Q} -\ln Q - \beta\frac{\pd{\beta}{Q}}{Q}}Q \\
    &= -\int \dd^4\mu\; (2|\alpha|^2+1)(\cj{\beta}\pd{\cj{\beta}}{Q}-\beta\pd{\beta}{Q})\\
    &=-\left[(2|\alpha|^2+1)\cj{\beta}Q-\text{c.c.}\right]_{-\infty}^{\infty} +\int \dd^4\mu\;(2|\alpha|^2+1)(Q - Q) = 0,
\end{split}
\end{equation}
the last term in the first line of Eq.~\eqref{eqn:X2_differential_operator} is structurally the same as the one above and hence it vanishes. The second contribution is given by,
\begin{equation}
    \begin{split}
        \Lambda_{x^2}^{(2)} &\propto \int \dd^4\mu\; (\cj{\alpha}^2\beta\pd{\cj{\beta}}{Q}-\alpha^2\cj{\beta}\pd{\beta}{Q})\ln Q = \left[Q\ln Q\cj{\alpha}^2\beta-\text{c.c.}\right]_{-\infty}^{\infty}-\int \dd^4\mu \; \cj{\alpha}^2\beta \pd{\cj{\beta}}{Q}-\alpha^2\cj{\beta} \pd{\beta}{Q}\\
        &=-\left[\cj{\alpha}^2\beta Q - \text{c.c.}\right]_{-\infty}^{\infty}+\int \dd^4\mu \; (\pd{\cj{\beta}}{[\cj{\alpha}^2\beta]}-\pd{\beta}{[\alpha}^2\beta])Q = 0,
    \end{split}
\end{equation}
and the same result holds for the last contribution of the second line of Eq.~\eqref{eqn:X2_differential_operator}. The first non-vanishing contribution comes from the third line,
\begin{equation}
    \begin{split}
        \Lambda_{x^2}^{(3)} &\propto \int \dd^4\mu\;\ln Q(\cj{\alpha}\cj{\beta}\pd{\cj{\alpha}}{\pd{\cj{\beta}}{Q}}-\alpha\beta\pd{\alpha}{\pd{\beta}{Q}}) = -\int \dd^4\mu \bigpar{\cj{\alpha}\frac{\pd{\cj{\alpha}}{Q}}{Q}\cj{\beta}\frac{\pd{\cj{\beta}}{Q}}{Q}-\text{c.c.}}Q,
    \end{split}
\end{equation}
and finally,
\begin{equation}
\label{4_th_line_ent_prod}
    \begin{split}
        \Lambda_{x^2}^{(4)} &\propto \int \dd^4\mu\;\ln Q(\cj{\alpha}^2\dpd{\cj{\beta}}{Q}-\alpha^2\dpd{\beta}{Q}) = \left[\cj{\alpha}\ln Q\pd{\cj{\beta}}{Q}-\text{c.c.}\right]_{-\infty}^{\infty} - \int \dd^4\mu \; \cj{\alpha}^2\frac{(\pd{\cj{\beta}}{Q})^2}{Q} - \text{c.c.}\\
        &= -\int \dd^4\mu \; \bigpar{\cj{\alpha}\frac{\pd{\cj{\beta}}{Q}}{Q}}^2 Q - \text{c.c.}
    \end{split}
\end{equation}
the second term of $\mathcal{J}_{x^2}^{(4)}$ is structurally the same as the first, it suffices to substitute $\alpha \rightarrow \beta$ in Eq.~\eqref{4_th_line_ent_prod}. Hence, we obtain that the non-vanishing contribution to $\Lambda_{x^2}$ is,
\begin{equation}
    \label{eqn:LambdaXTotal}
   \Lambda_{x^2}= \frac{i\gamma_x}{8j}\int \dd^4\mu \; \left[\bigpar{\cj{\alpha}\frac{\pd{\cj{\beta}}{Q}}{Q}}^2+2\cj{\alpha}\cj{\beta}\frac{\pd{\cj{\alpha}}{Q}}{Q}\frac{\pd{\cj{\beta}}{Q}}{Q} +\bigpar{\cj{\beta}\frac{\pd{\cj{\alpha}}{Q}}{Q}}^2 - \text{c.c.}\right]Q= \frac{i\gamma_x}{8j}\int \dd^4\mu \;\left[\bigpar{\cj{\alpha}\frac{\pd{\cj{\beta}}{Q}}{Q}+\cj{\beta}\frac{\pd{\cj{\alpha}}{Q}}{Q}}^2 - \text{c.c.}\right]Q.
\end{equation}
The contribution due to $\Lambda_z$ vanishes because it has exactly the same structure as $\Lambda_{x}^{(1)}$ in Eq.~\eqref{eqn:LambdaX1}. Now, Eq. \eqref{eqn:LambdaXTotal} can be written as
\begin{equation}
\begin{split}
    \Lambda_{x^2} &= \frac{\gamma_x}{4j}\Im{\left\langle\bigpar{\cj{\alpha}\frac{\pd{\cj{\beta}}{Q}}{Q}+\cj{\beta}\frac{\pd{\cj{\alpha}}{Q}}{Q}}^2\right\rangle}\\
    &=-\frac{\gamma_x}{4j}\Im{\left\langle\bigpar{\alpha\frac{\pd{\beta}{Q}}{Q}+\beta\frac{\pd{\alpha}{Q}}{Q}}^2\right\rangle},
\end{split}
\end{equation}
thus leading us to the following expression for the entropy rate
\begin{equation}
    \frac{\dd S_{Q}}{\dd t} = - \Lambda_{x^2}= \frac{\gamma_x}{4j}\Im{\left\langle\bigpar{\alpha\frac{\pd{\beta}{Q}}{Q}+\beta\frac{\pd{\alpha}{Q}}{Q}}^2\right\rangle}.
\end{equation}
In this equation, $\mean{...}$ means an average over the phase-space.

For a fixed the value of $j$, we can go back to the polar representation ($\theta$,$\phi$) using the following relations~\cite{Takahashi1975},
\begin{eqnarray}
    \alpha\pd{\beta}{}&=e^{-i\phi}\cos^2{\frac{\theta}{2}}\left[2j\tan{\frac{\theta}{2}}+\pd{\theta}{}-i\frac{\pd{\phi}{}}{\sin{\theta}}\right]\\[0.2cm]
    \beta\pd{\alpha}{}&=e^{i\phi}\sin^2{\frac{\theta}{2}}\left[2j\cot{\frac{\theta}{2}}-\pd{\theta}{}+i\frac{\pd{\phi}{}}{\sin{\theta}}\right]
\end{eqnarray}
since $S_Q(\mu) = - \int \dd^4\mu\;\ln{Q(\mu)}Q(\mu) \rightarrow S_Q(\Omega) = -(2j+1)/4\pi \int \dd^4\Omega\;\ln{Q(\Omega)}Q(\Omega) $ which gives us the following expression for the entropy production in polar coordinates,
\begin{equation}
    \frac{\dd S_{\Omega}}{\dd t} = \frac{2j+1}{16\pi j}\gamma_x\Im{\int d\Omega\; \frac{1}{Q(\Omega)}\bigpar{e^{-i\phi}\cos^2{\frac{\theta}{2}}\left[2j\tan{\frac{\theta}{2}}+\pd{\theta}{}-i\frac{\pd{\phi}{}}{\sin{\theta}}\right]Q(\Omega)+Q(\Omega)\sin^2{\frac{\theta}{2}}\left[2j\cot{\frac{\theta}{2}}-\pd{\theta}{}+i\frac{\pd{\phi}{}}{\sin{\theta}}\right]Q(\Omega)}^{2}}
\end{equation}
\end{widetext}

%
%

\section{Numerical studies on the LMG model}
\label{app:num}

In studying the DPT of the LMG model, care must be taken with the fact that the ground-state is two-fold degenerate. We therefore define an echo for each ground-state, $\mathcal{L}_{\alpha} = \left\vert \langle\psi_{0}^{\alpha}\vert\psi_{t}^{\alpha}\rangle \right\vert^{2}$ and consider only the \emph{net} rate 
\begin{equation}
r_{s}(t) = -\frac{1}{N}\log\left[\sum_{\alpha=1}^{g}\mathcal{L}_{\alpha}\right],
\label{eq:rate}
\end{equation}
which therefore captures the total return probability. 
However, as shown \cite{Heyl2018},  in the thermodynamic limit this converges to the minimum among all the contributions 
\begin{equation}
r_{m} \equiv \lim_{N\rightarrow \infty} r(t) = -\frac{1}{N}\log\left[\min_{\alpha}\mathcal{L}_{\alpha}\right].
\label{eq:rate_limit}
\end{equation}
It is for this reason that in the main text it sufficed to consider only the rate starting from a single ground-state. 

In this Appendix we describe several numerical studies of the LMG model. Specifically, we consider four distinct quench processes, from $h_0 = 0$ to $h = 0.1$, $0.8$, $1.0$ and $1.6$. Note that the first two quenches occur before the quantum critical point while the last one occurs after.

\subsection{The dynamical quantum phase transition}

\begin{center}
    \begin{figure}[H]
        \centering
        \includegraphics[width=0.5\textwidth]{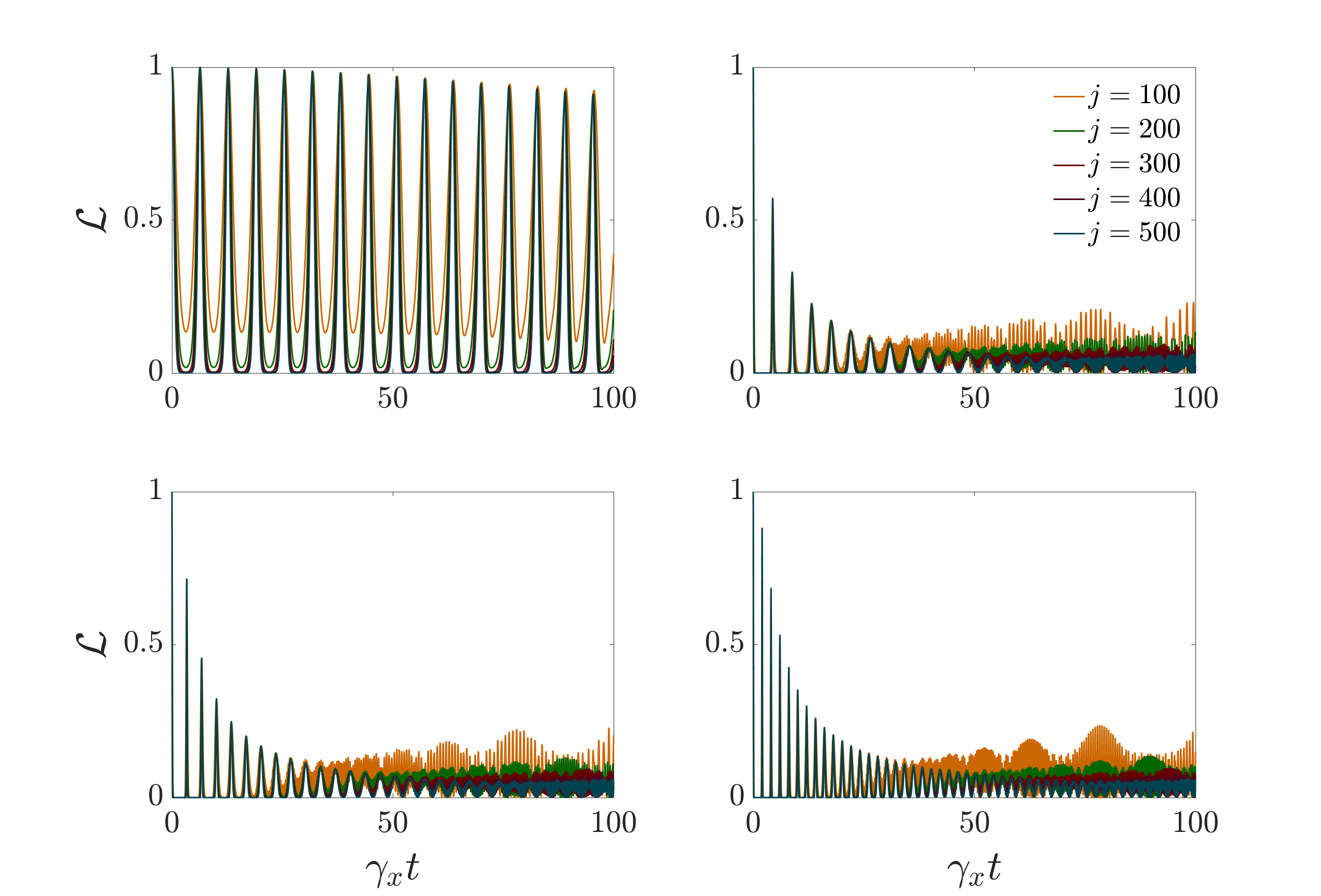}
        \caption{Loschmidt echo for the first ground state. From top to bottom and from left to right we consider  quenches from $h_0 = 0$ to $h = 0.1$, $0.8$, $1.0$ and $1.6$.}
        \label{app:fig:echo01}
    \end{figure}
\end{center}

\begin{center}
    \begin{figure}[H]
        \centering
        \includegraphics[width=0.5\textwidth]{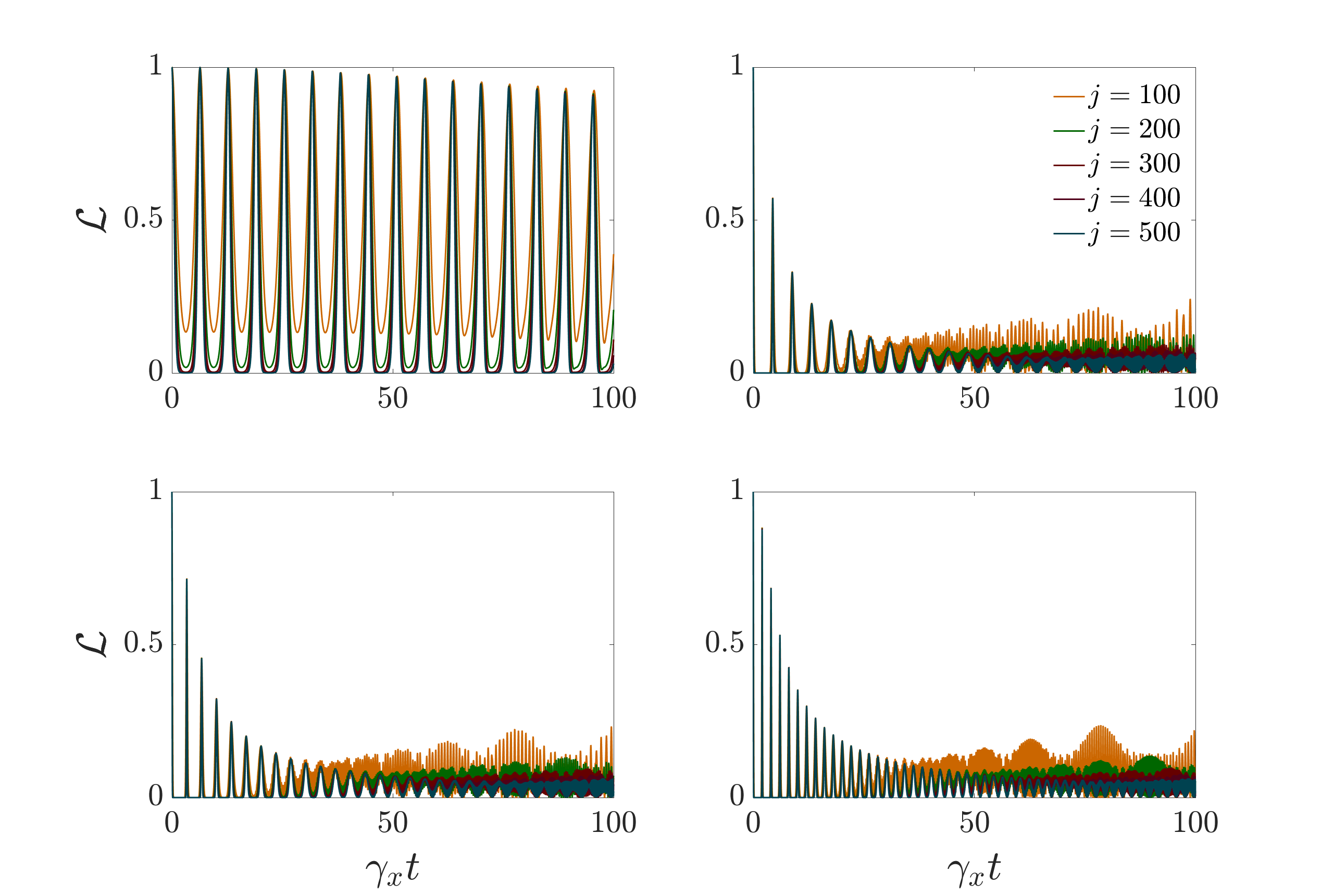}
        \caption{Loschmidt echo for the second ground state. From top to bottom and from left to right we consider quenches from $h_0 = 0$ to $h = 0.1$, $0.8$, $1.0$ and $1.6$.}
        \label{app:fig:echo02}
    \end{figure}
\end{center}

We start by showing the Loschmidt echo for the two ground states of the system in Figs. \ref{app:fig:echo01} and \ref{app:fig:echo02}. As we can see from these figures, the behaviour of $\mathcal{L}$ regarding the dynamical quantum phase transition is independent of the considered initial state. However, it does depend on the quench amplitude and show no sensible difference at the quantum critical point $h=1$.

\begin{center}
    \begin{figure}[H]
        \centering
        \includegraphics[width=0.5\textwidth]{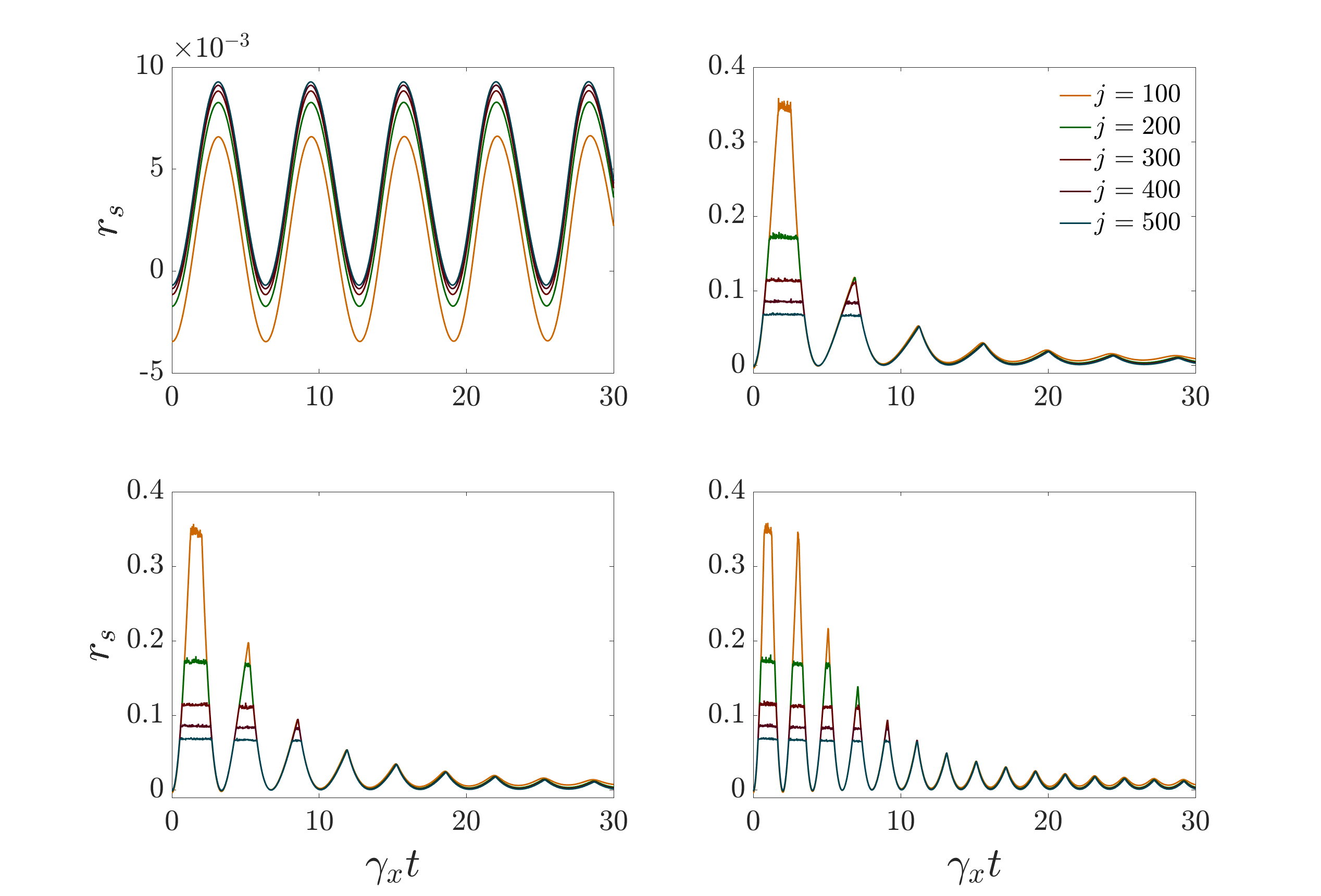}
        \caption{Rate function for the total return probability. From top to bottom and from left to right we consider quenches from $h_0 = 0$ to $h = 0.1$, $0.8$, $1.0$ and $1.6$.}
        \label{app:fig:rate_s}
    \end{figure}
\end{center}

\begin{center}
    \begin{figure}[H]
        \centering
        \includegraphics[width=0.5\textwidth]{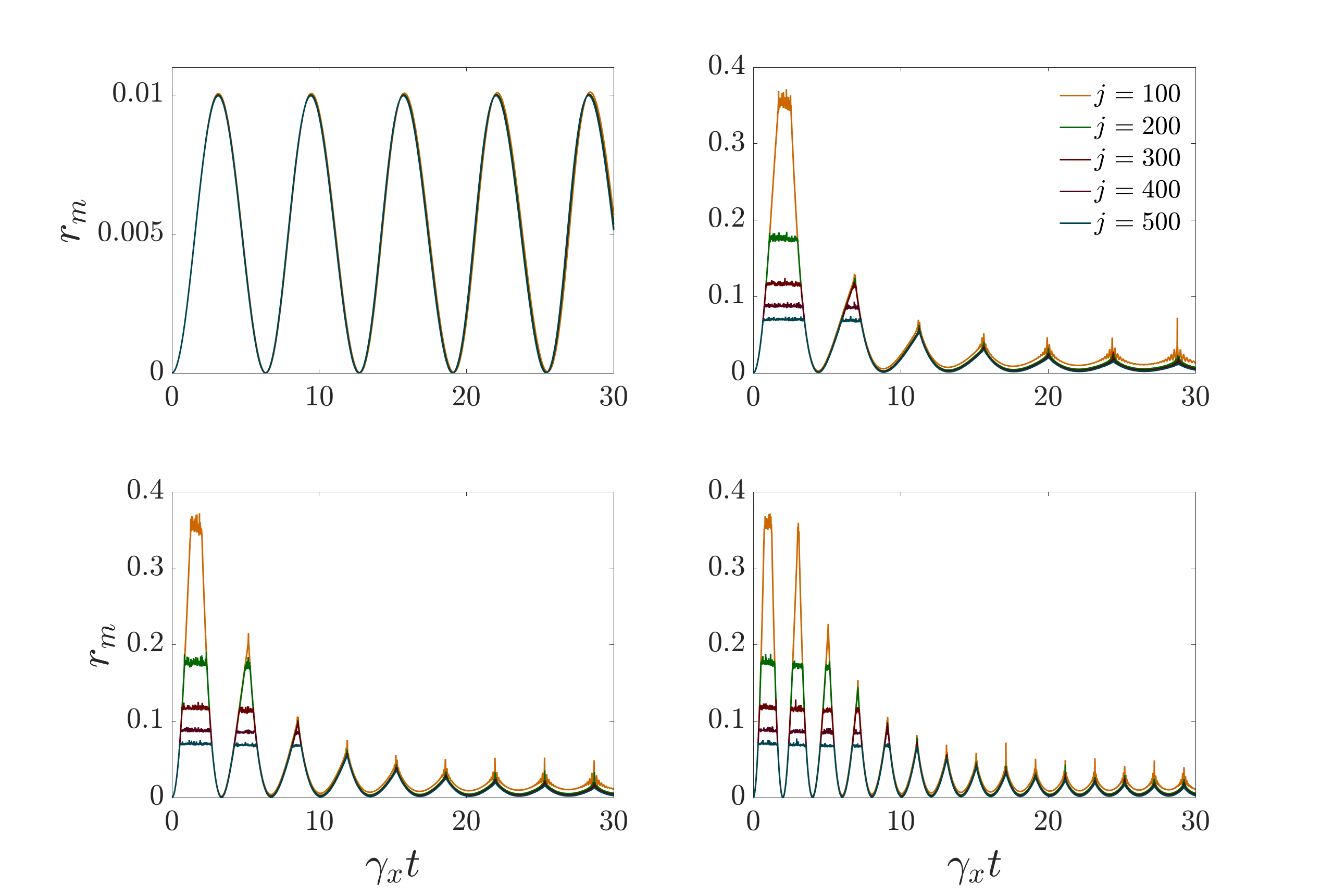}
        \caption{Rate function for the minimum return probability. From top to bottom and from left to right we consider quenches from $h_0 = 0$ to $h = 0.1$, $0.8$, $1.0$ and $1.6$.}
        \label{app:fig:rate_m}
    \end{figure}
\end{center}

\begin{center}
    \begin{figure}[H]
        \centering
        \includegraphics[width=0.5\textwidth]{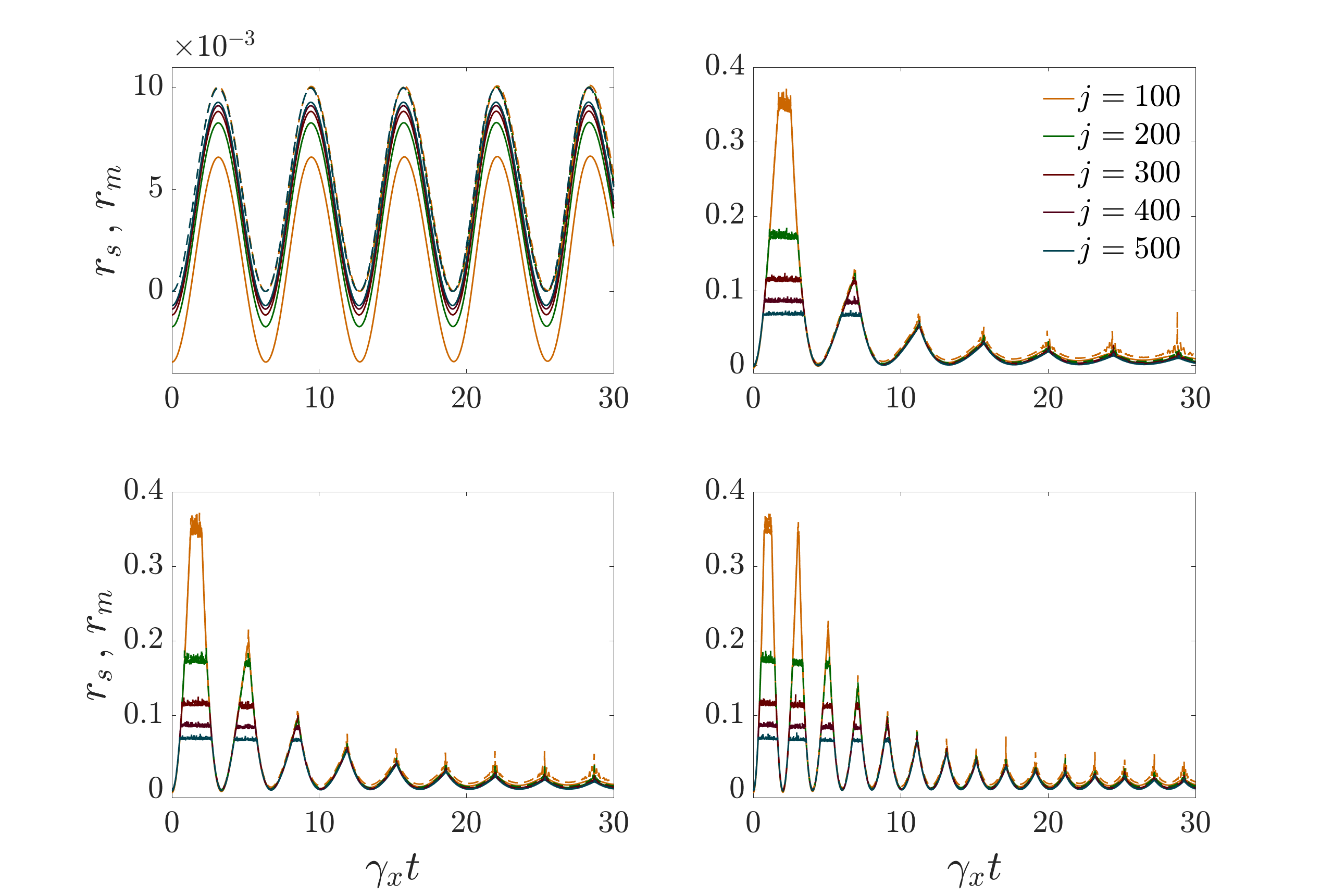}
        \caption{Comparison between $r$ (solid lines) and $r_{m}$ (dashed lines). From top to bottom and from left to right we consider quenches from $h_0 = 0$ to $h = 0.1$, $0.8$, $1.0$ and $1.6$.}
        \label{app:fig:comp_rates}
    \end{figure}
\end{center}

To make this point more clear, in Figs. \ref{app:fig:rate_s} and \ref{app:fig:rate_m} we show the rate functions $r$ and $r_{m}$, respectively. It is clear that the dynamical quantum phase transition depends on the quench amplitude and on the size of the angular momentum. In order to see that $r_{s}$ converges to $r_{m}$ when we approach the thermodynamic limit, in Fig. \ref{app:fig:comp_rates} we show both quantities for distinct values of the angular momentum.  

\subsection{Entropy production}

\begin{center}
    \begin{figure}[ht!]
        \centering
        \includegraphics[width=0.5\textwidth]{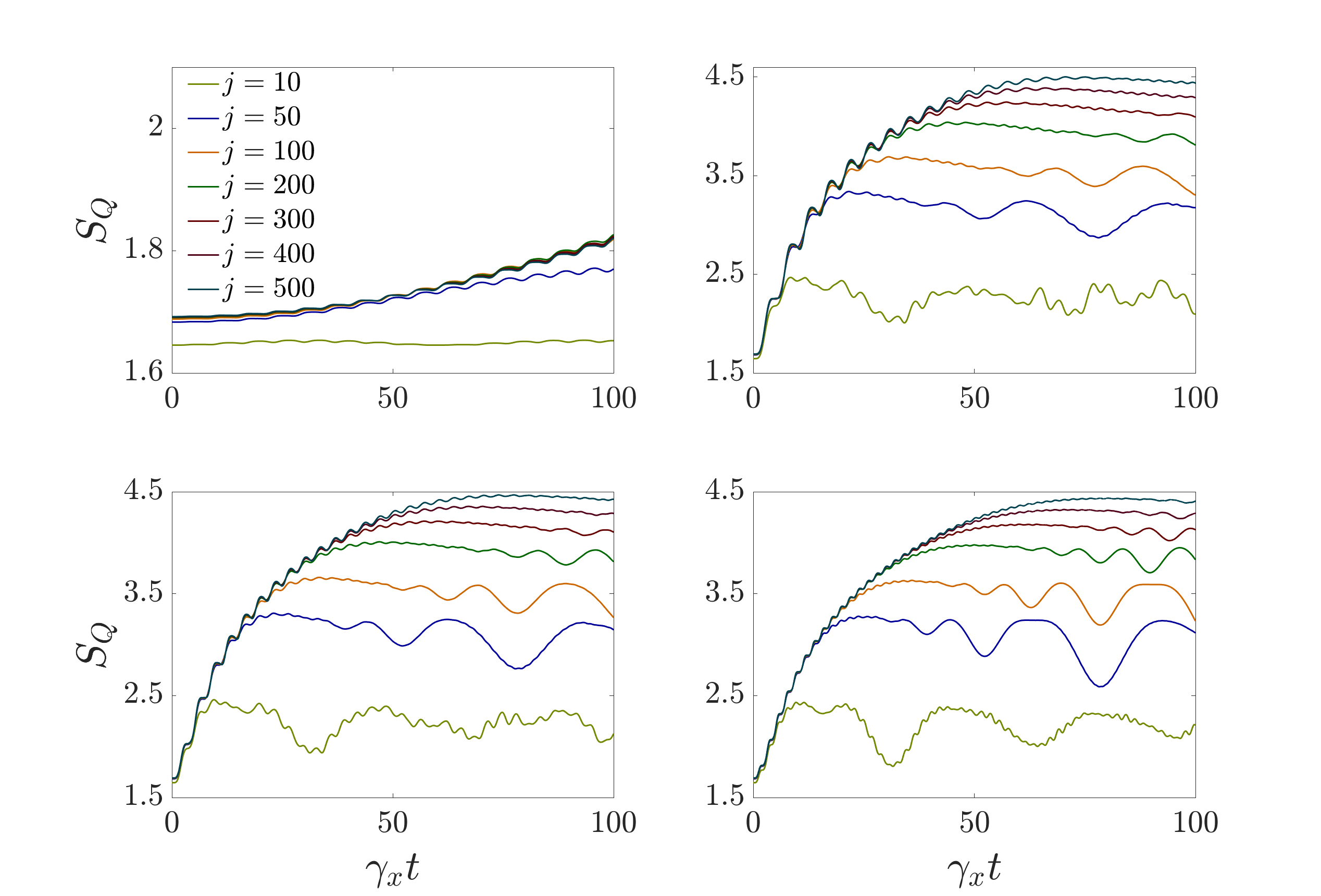}
        \caption{Entropy for the first ground state. From top to bottom and from left to right we consider quenches from $h_0 = 0$ to $h = 0.1$, $0.8$, $1.0$ and $1.6$.}
        \label{app:fig:entro01}
    \end{figure}
\end{center}

\begin{center}
    \begin{figure}[ht!]
        \centering
        \includegraphics[width=0.5\textwidth]{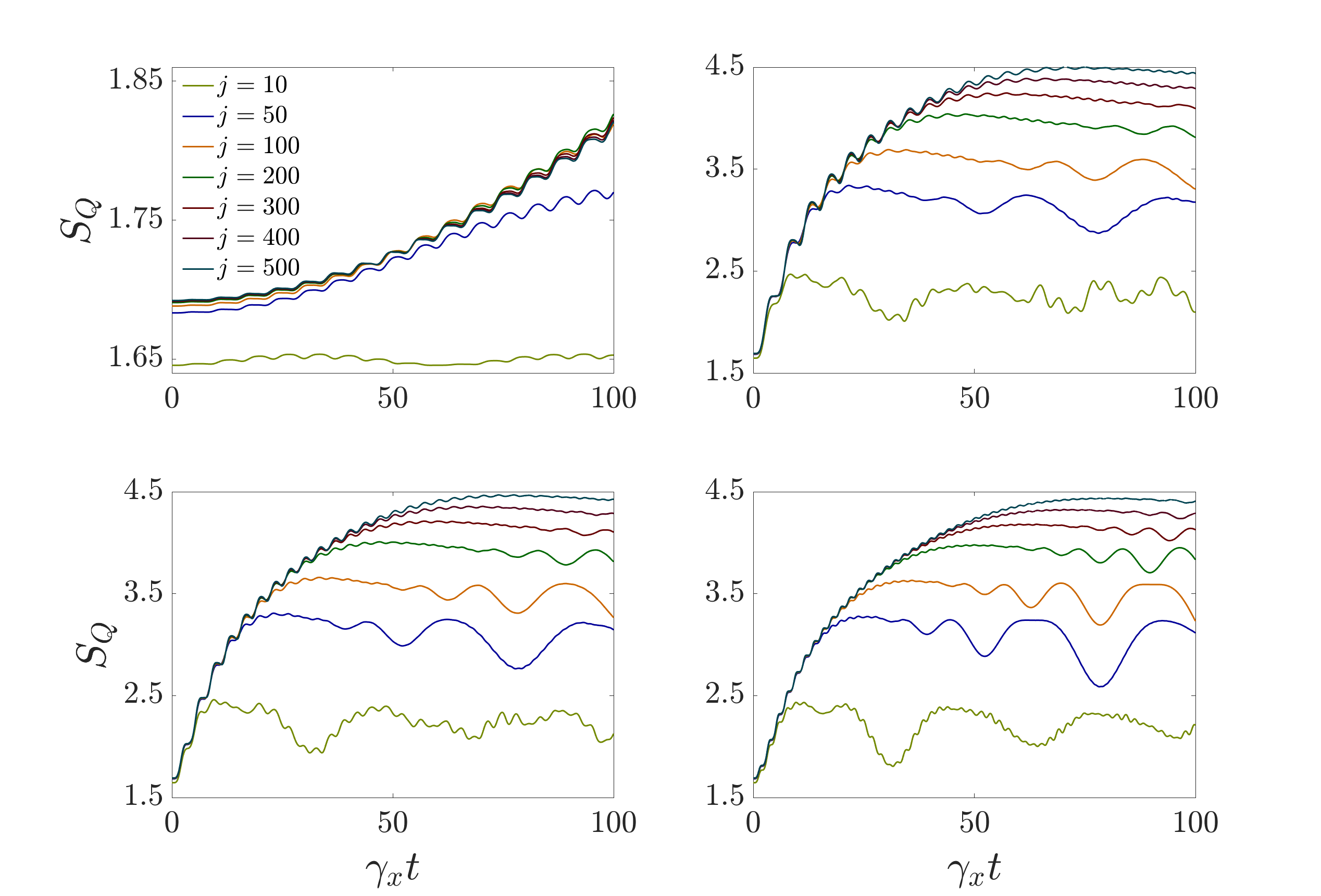}
        \caption{Entropy for the second ground state. From top to bottom and from left to right we consider quenches from $h_0 = 0$ to $h = 0.1$, $0.8$, $1.0$ and $1.6$.}
        \label{app:fig:entro02}
    \end{figure}
\end{center}

\begin{center}
    \begin{figure}[ht!]
        \centering
        \includegraphics[width=0.5\textwidth]{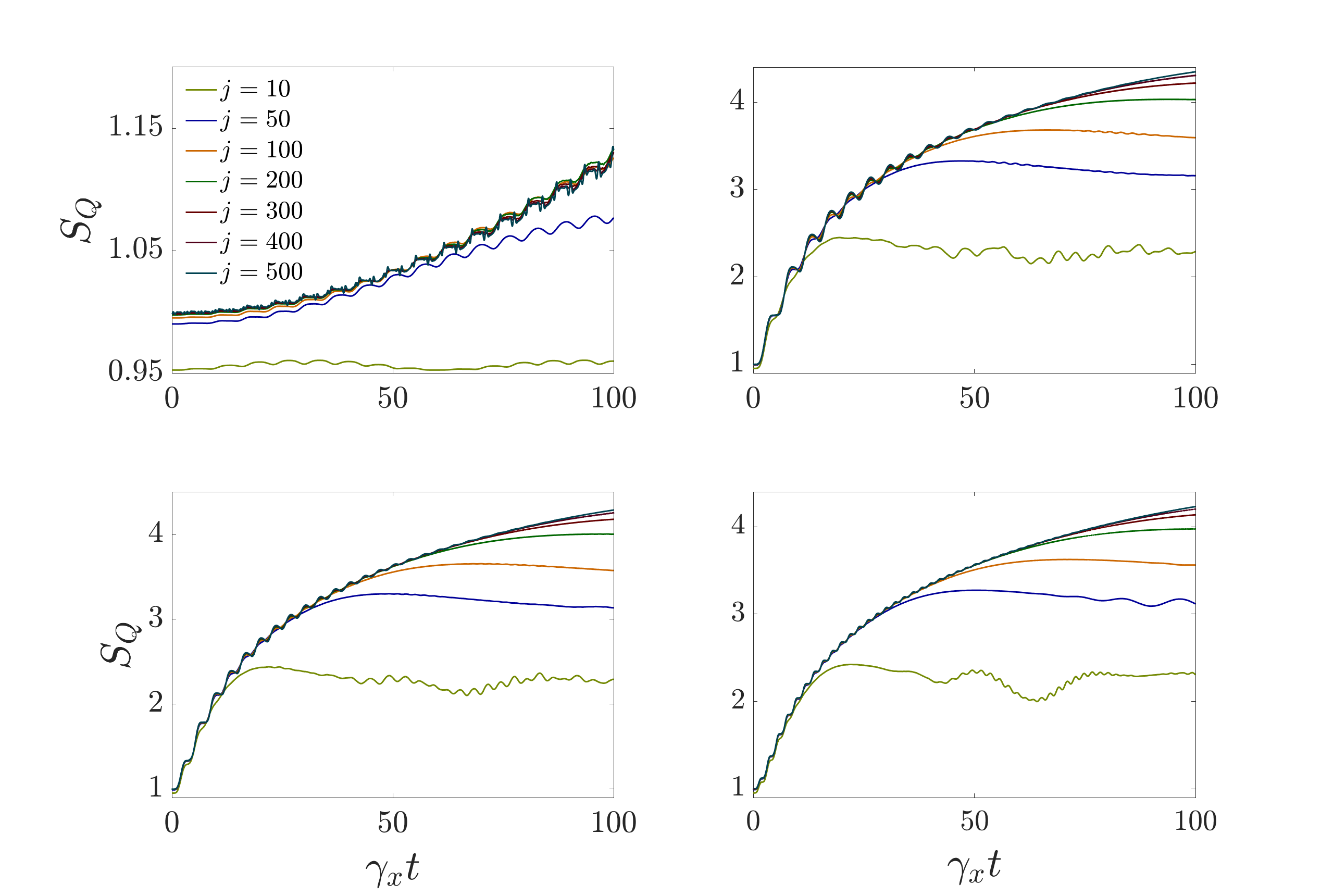}
        \caption{Entropy for the superposition of both ground states. From top to bottom and from left to right we consider quenches from $h_0 = 0$ to $h = 0.1$, $0.8$, $1.0$ and $1.6$.}
        \label{app:fig:entros}
    \end{figure}
\end{center}

Figures \ref{app:fig:entro01}, \ref{app:fig:entro02} and \ref{app:fig:entros} show the dynamical behaviour of the entropy for three distinct states, the two ground states and the equal superposition of these states. We can observe the same qualitative behaviour for all the considered quenches, except for the smaller one. The entropy approaches a maximum as we approach the thermodynamic limit. The oscillation pattern presented in these plots signal the dynamical quantum phase transitions.

\begin{center}
    \begin{figure}[ht!]
        \centering
        \includegraphics[width=0.5\textwidth]{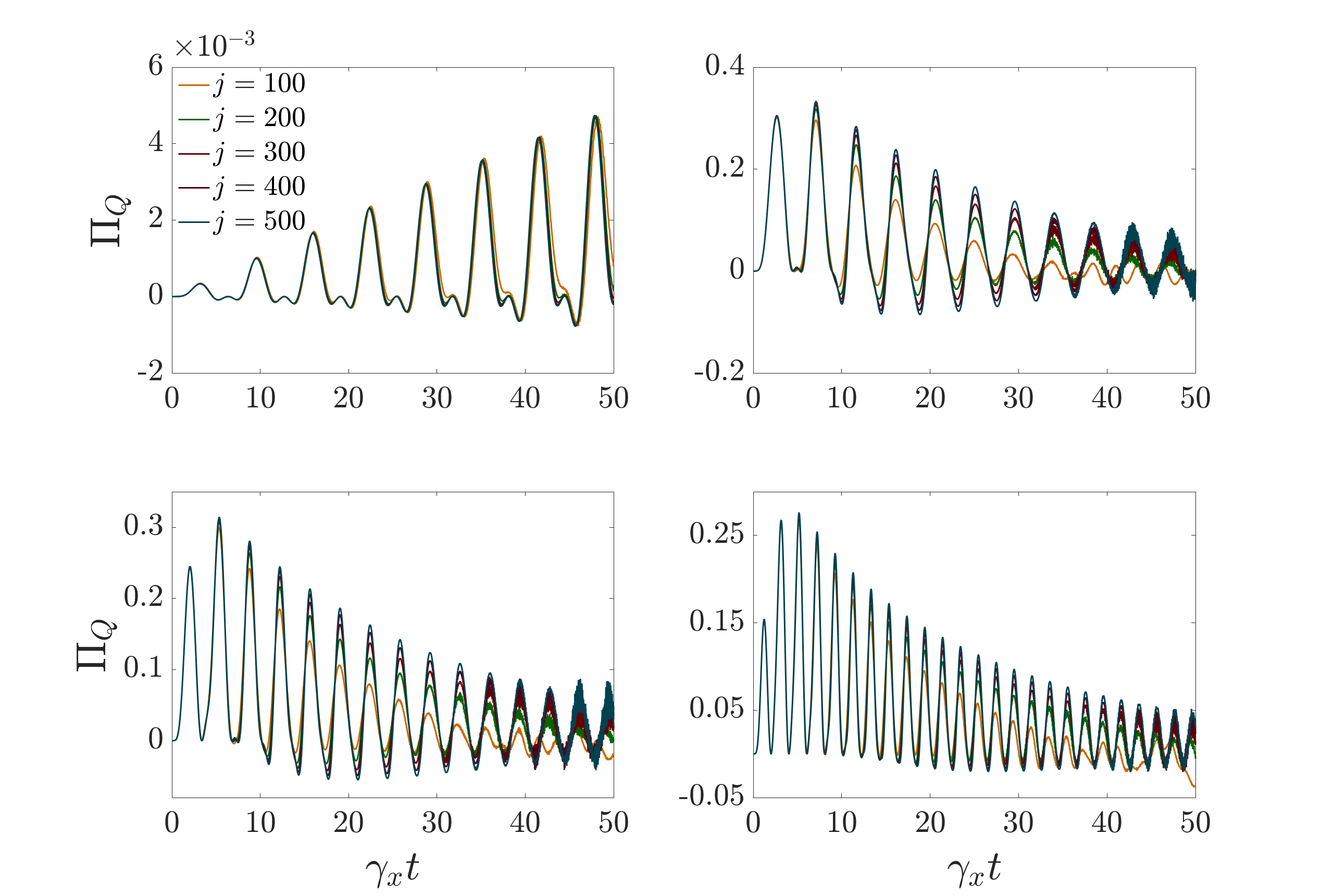}
        \caption{Entropy production rate for the first ground state. From top to bottom and from left to right we consider quenches from $h_0 = 0$ to $h = 0.1$, $0.8$, $1.0$ and $1.6$.}
        \label{app:fig:entro_rate01}
    \end{figure}
\end{center}

\begin{center}
    \begin{figure}[ht!]
        \centering
        \includegraphics[width=0.5\textwidth]{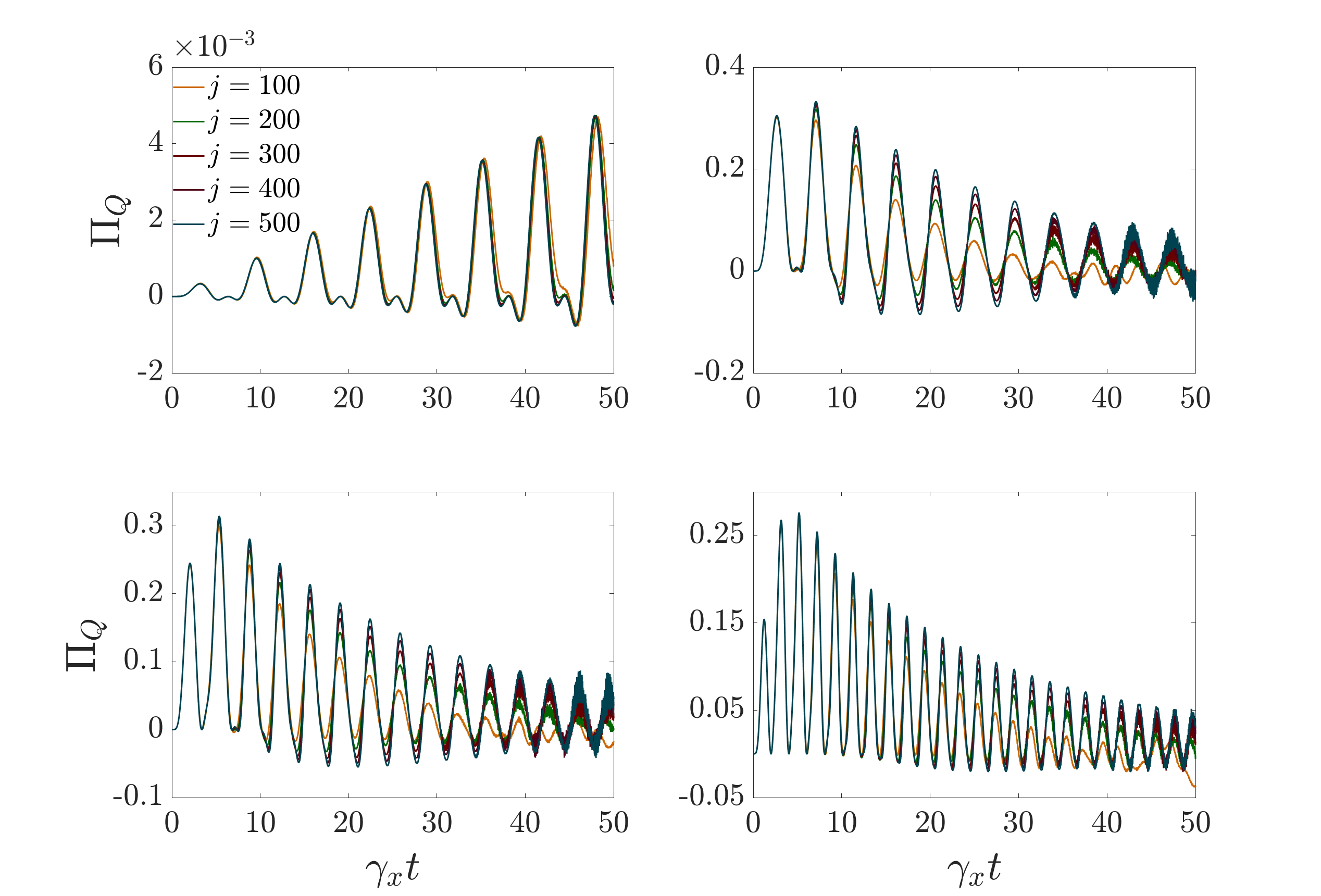}
        \caption{Entropy production rate for the second ground state. From top to bottom and from left to right we consider quenches from $h_0 = 0$ to $h = 0.1$, $0.8$, $1.0$ and $1.6$.}
        \label{app:fig:entro_rate02}
    \end{figure}
\end{center}

\begin{center}
    \begin{figure}[H]
        \centering
        \includegraphics[width=0.5\textwidth]{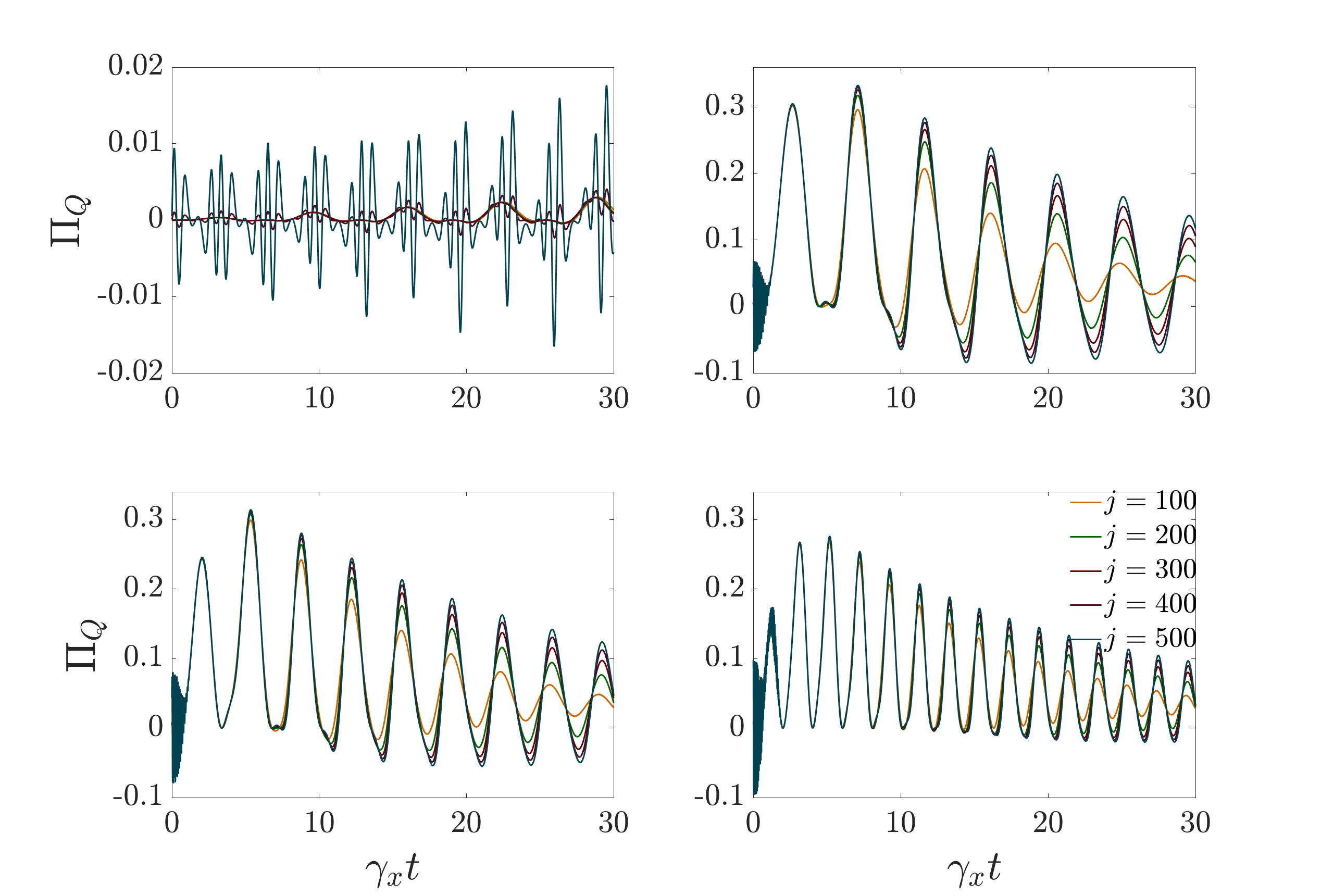}
        \caption{Entropy production rate for the superposition of both ground states. From top to bottom and from left to right we consider quenches from $h_0 = 0$ to $h = 0.1$, $0.8$, $1.0$ and $1.6$.}
        \label{app:fig:entro_rates}
    \end{figure}
\end{center}

In order to show this fact, we consider the entropy production rate, given by the time derivative of $S_{Q}$. This is shown in Figs. \ref{app:fig:entro_rate01}, \ref{app:fig:entro_rate02} and \ref{app:fig:entro_rates} for the same three states addressed in the case of the entropy. Again, except for the small quench, the qualitative behaviour of this quantity presents several maximums and we show in the main text and in the next subsection that such maximums are related to the dynamical quantum phase transitions.

\subsection{Entropy production rate and the rate function}

\begin{center}
    \begin{figure}[H]
        \centering
        \includegraphics[width=0.5\textwidth]{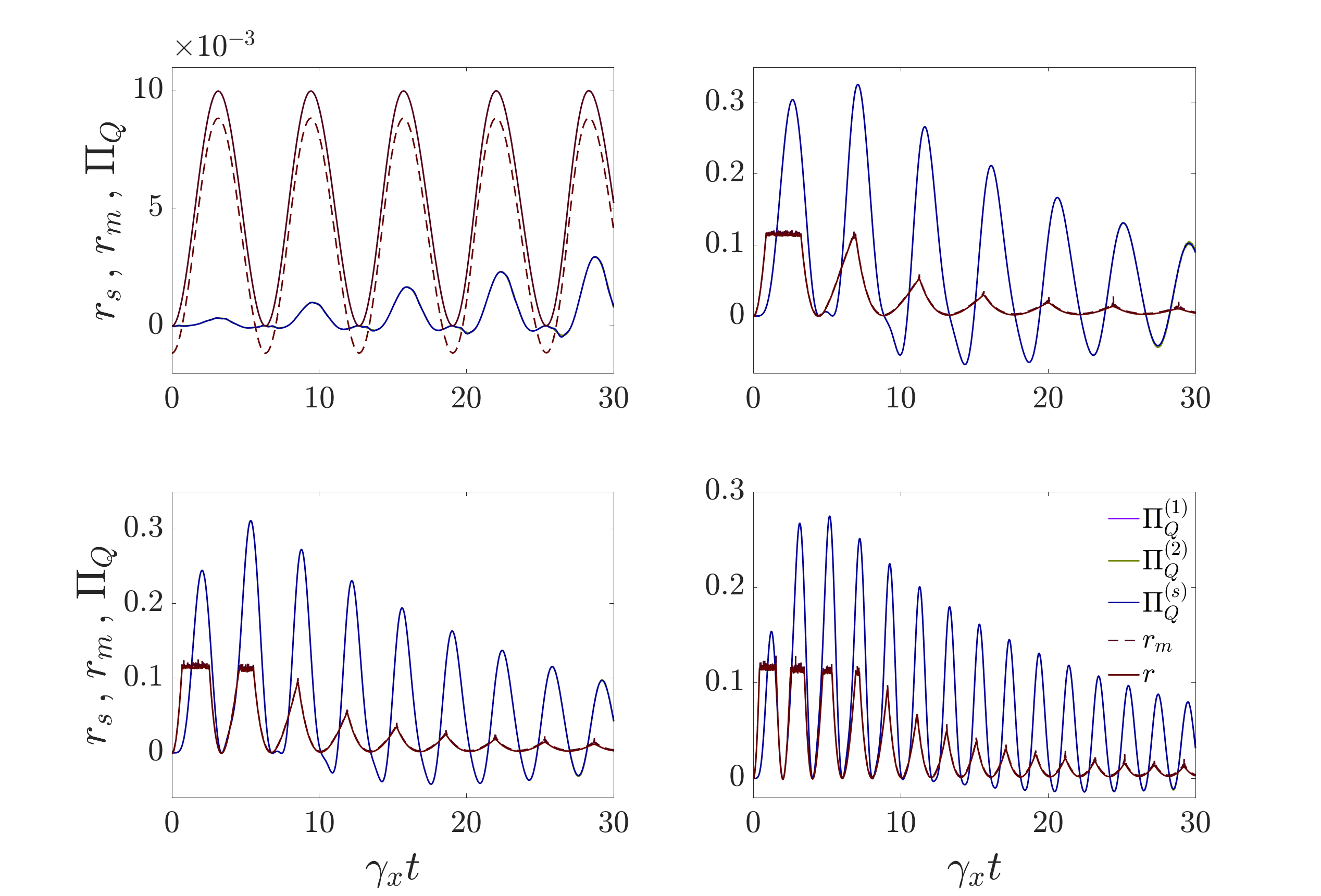}
        \caption{Comparison between $r$ and $r_{m}$. From top to bottom and from left to right we consider quenches from $h_0 = 0$ to $h = 0.1$, $0.8$, $1.0$ and $1.6$.}
        \label{app:fig:comp_rate_entro}
    \end{figure}
\end{center}

Finally, we address here the main message of the present article. Figure \ref{app:fig:comp_rate_entro} shows the comparison between the entropy production rate and the rate function $r_{m}$ for a fixed value of the angular momentum ($j=500$). Except for the case of small quench, where we actually do not have a quantum phase transition since $r_{m}$ is analytical in time, the other considered quenches clearly shows that the entropy production rate is able to sign the dynamical quantum phase transition, as discussed in the main text.

\begin{center}
    \begin{figure}[H]
        \centering
        \includegraphics[width=0.5\textwidth]{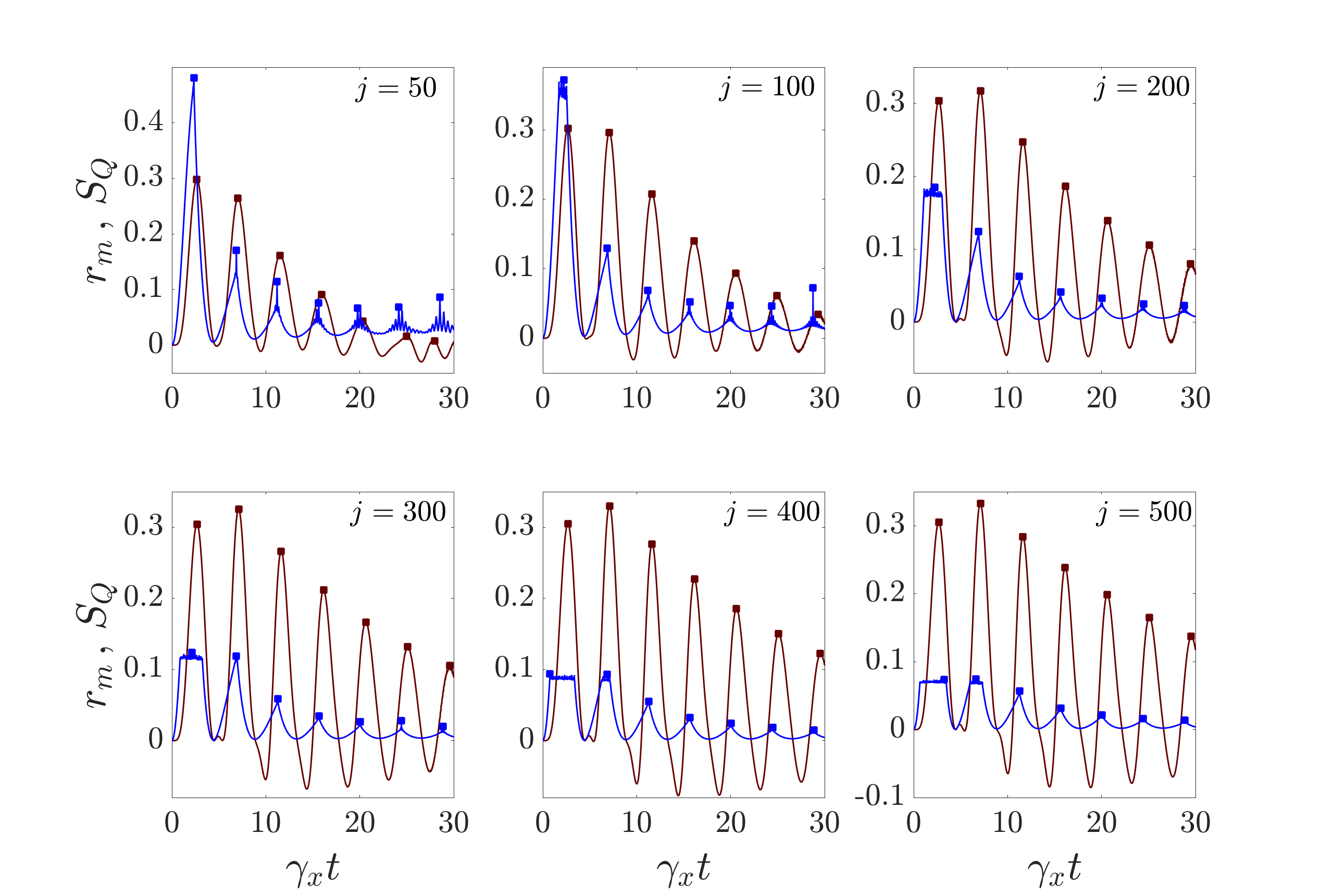}
        \caption{Comparison between the maximums of the entropy production rate and the non-analytical points of the rate function for the quench $h_0 = 0$ to $h=0.8$.}
        \label{app:fig:critical}
    \end{figure}
\end{center}

This relation can be deeply stated if we consider the times at which entropy production rate shows a maximum and the critical times, where we have a dynamical quantum phase transition. This is done in Fig. \ref{app:fig:critical}, where a clear linear behaviour emerges, thus supporting our claims in the main text. The points highlighted in this figure are the ones shown in Fig. \ref{fig:critical} of the main text. 

%
%

\end{document}